\documentclass[preprint,10pt]{elsarticle}
\usepackage{graphicx}
\usepackage{color}
\newcommand{\rmexp}{\mathrm{exp}}
\newcommand{\calN}{\mathcal{N}}
\newcommand{\rmor}{\mathrm{or}}
\newcommand{\rmtot}{\mathrm{tot}}
\newcommand{\rmc}{\mathrm{c}}
\newcommand{\rme}{\mathrm{e}}
\newcommand{\rmi}{\mathrm{i}}
\newcommand{\rmR}{\mathrm{R}}
\newcommand{\rmg}{\mathrm{g}}

\newcommand{\rmwith}{\mathrm{with}}
\newcommand{\rmand}{\mathrm{and}}

\begin{document}
\title{Time-dependent pointer states and determination of the preferred basis of measurement}
\author{Hoofar Daneshvar \corref{cor1}}
\ead{hoofar@uwindsor.ca}
\author{G.W.F. Drake}
\ead{GDrake@uwindsor.ca}
\address{Department of Physics, University of Windsor, Windsor ON, N9B 3P4, Canada}
\cortext[cor1]{Corresponding author. Address: Department of Physics, University of Windsor, Windsor ON, N9B 3P4, Canada. Tel.: +1 519 2533000 2689.}
\begin{abstract}
We present a general analytic method for evaluating the generally time-dependent pointer states of a subsystem, which are defined by their capability not to entangle with the states of another subsystem. In this way, we show how in practice the global state of the system and the
environment may evolve into a diagonal state (i.e.\ the Von Neumann scheme of measurement may be realized) as a result of the natural evolution of the total
composite system. We explore the conditions under which the pointer states of the system become independent of time; so that a preferred basis of measurement can be realized. As we show, these conditions include the so-called quantum limit of decoherence and the so-called quantum measurement limit; as well as some other specific conditions which are discussed in the paper. We relate the mathematical conditions for having time-independent pointer states to some classes of possible symmetries in the Hamiltonian of the total composite system. Indeed, our theory would serve as a generalization of the existing theory for determination of the preferred basis of measurement. By exploiting this new theory we can obtain those regimes of the parameter space for a given total Hamiltonian defining our system-environment model for which a preferred basis of measurement can be realized. Moreover, we can predict the corresponding preferred basis of measurement for each regime. Our theory contains a powerful method for obtaining the time-dependent pointer states of the system and the environment in most of the other regimes where the pointer states of the system are time-dependent and a preferred basis of measurement cannot be realized at all. This ability to obtain time-dependent pointer states is specifically important in decoherence studies; as these pointer states, although they evolve with time and cannot represent the preferred basis of measurement, they correspond to those initial conditions for the state of the system and the environment for which we can have longer decoherence times.
\end{abstract}
\maketitle

\section{Introduction}

This is the first paper in the series of papers where we discuss the pointer states of measurement and Born's rule, $p_{k}=|\psi_{k}|^{2}$, for quantum probabilities. We refer to this paper as paper $\sf I$ in this series of papers. This paper is organized as follows:

In section 2 (which is our introductory review and discussion section to better identify what the problem is) after reviewing the orthodox theory for determination of the pointer states of measurement we discuss an important restriction of this theory regarding the so-called \emph{commutativity criterion} for determination of the pointer states. In fact, we will show that the pointer states of the system and the environment, which appear in the diagonal state of the total composite system
\begin{equation}\label{1}
    |\psi_{S\cal E}\rangle=\sum_{k=1}^{N}
    \alpha_{k}|s_{k}\rangle|\varepsilon_{k}\rangle
\end{equation}
after premeasurement by the environment \footnote{In this paper whenever we talk about the environment it refers to all the subsystems which are outside our system of interest and hence it can include the apparatus as well.}, generally are time-dependent and the commutativity criterion, first introduced by Zurek [1-3], although is sufficient in order to assure the requirement of having a faithful measurement, it is too restrictive and generally does not hold valid for the pointer states of measurement in all situations. In other words, it is not the \emph{minimal} condition for having a faithful measurement. We will show that the commutativity criterion can be valid \emph{only} in certain regimes and under the specific conditions which the pointer states of the system \footnote{The pointer states of a system are characterized by their ability not to entangle with the states of the environment (i.e.\ the requirement of faithful measurement) and appear in the diagonal state of the total composite system after premeasurement by the environment. As we elaborately describe in this paper, generally we should
distinguish between the set of pointer states in equation (1) and the preferred basis of measurement; mainly because of the fact that the pointer states of a subsystem generally are time-dependent and a preferred basis of measurement does not exist, unless under the specific conditions (discussed in section 4) which the pointer states of measurement become time-independent. Moreover, the pointer states of a system necessarily are not orthonormal amongst themselves at all times. Therefore, necessarily they cannot form a basis for the Hilbert space of the system at all times.} are independent of time.

In section 2 we will also very briefly review some of the other predictability criteria which have been exploited to obtain the pointer states of measurement. We will discuss some of the restrictions and difficulties which one should expect while using these methods.

In section 3 we present a method in order to calculate the (generally) time-dependent pointer states of the system and the environment for an arbitrary total Hamiltonian defining the system-environment model and in section 4, using this method, we will exactly discuss under which conditions can we have time-independent pointer states; and also how we can predict the preferred basis of measurement in each of the corresponding regimes. As we will see, only under specific conditions time-independent pointer states can be realized; therefore a preferred basis of measurement necessarily does not exist in an arbitrary regime.

In section 5 we will discuss the significance of our theory more elaborately and will conclude.

\section{Review and discussion: identifying the problem}
\subsection{The Schmidt decomposition}

Before proceeding to the question of the pointer states of measurement we would talk about the old \emph{Schmidt Decomposition Theorem} \cite{Schmidt} which states that an arbitrary pure state $|\Psi\rangle$ of the composite system $\cal AB$, made up of two subsystems $\cal A$ and $\cal B$ endowed with Hilbert spaces $\cal H_{\cal A}$ and $\cal H_{\cal B}$, can always be written in a diagonal form
\begin{equation}\label{2}
|\Psi\rangle=\sum_{i}\lambda_{i}\ |a_{i}\rangle\ |b_{i}\rangle,
\end{equation}
where the Schmidt states $|a_{i}\rangle$ and $|b_{i}\rangle$ are orthonormal amongst themselves and form the so-called \emph{Schmidt bases} of $\cal H_{\cal A}$ and $\cal H_{\cal B}$ respectively, and the expansion coefficients $\lambda_{i}$ generally are some complex numbers fulfilling $\sum_{i}|\lambda_{i}|^{2}=1$. Moreover, it is shown that this decomposition is unique if and only if the expansion coefficients $\lambda_{i}$ are all different from one another. Note that the above statement basically refers to the fact that in describing the total state of a composite system in a diagonal form with orthonormal basis states generally there is a \emph{basis ambiguity}, as the diagonal decomposition is not always unique.

We also note that the reduced density matrices $\hat{\rho}_{\cal A}$ and $\hat{\rho}_{\cal B}$ for the two subsystems which are obtained by tracing operation will be diagonal in the Schmidt bases $\{|a_{i}\rangle\}$ and $\{|b_{i}\rangle\}$, as one can easily verify, because of the fact that these states are orthogonal amongst themselves. Therefore, the Schmidt bases correspond to the orthonormal basis states which diagonalize the reduced density matrices.

\subsection{Premeasurement by the environment}

In order to define the pointer states of measurement and describe their exact distinction from the Schmidt bases and also from the preferred basis of measurement we need to describe what we
mean by a faithful measurement. Faithful measurement in the usual
sense concerns the requirement of a one-to-one correspondence
between the states of the system and the apparatus. Just in this
case, the state of the apparatus can be viewed as a reliable pointer
(indicator) for the state of the system. This requirement can
describe the Von Neumann scheme \cite{Neuman} for quantum measurement which states
that if the system starts out in a superposition of the basis states
$|s_{i}\rangle$,
\begin{equation}\label{3}
|\psi\rangle=\sum_{i} c_{i}|s_{i}\rangle
\end{equation}
then, the system-apparatus combination will evolve according to
\begin{equation}\label{4}
|\psi\rangle|a_{r}\rangle=(\sum_{i}
c_{i}|s_{i}\rangle)|a_{r}\rangle\rightarrow |\psi\rangle=\sum_{i}
c_{i}|s_{i}\rangle|a_{i}\rangle
\end{equation}
where $|a_{r}\rangle$ is the initial ``ready" state of the
apparatus. We note that the Von Neumann scheme described by equation
(4), which usually is also referred to as \emph{premeasurement}, can
be obtained just by assuming the requirement of the faithful measurement of
the state of the system initially prepared in the state
$|s_{i}\rangle$, i.e.\ the requirement that
$|s_{i}\rangle|a_{r}\rangle\rightarrow |s_{i}\rangle|a_{i}\rangle$
and the linearity of Schr\"{o}dinger's equation.

The states appearing in the righthand side of equation (4) are the pointer states of the two subsystems. The main characteristic of the pointer states of the system and the environment is their capability to maintain their individuality, as well as the one-to-one correspondence between themselves (the requirement of faithful measurement), during the interaction. This means that any states other than the pointer states of the system are subject to entanglement with the states of the environment so that they lose their individuality (meaning that one no longer can ascribe a well-defined state to the system alone, due to the entanglement with the environment) and one cannot consider a one-to-one correspondence between some well-defined states from the system and some states from the environment. However, the Von Neumann scheme assumes that the
measurement interaction is ideal in the sense that it does not
change the state of the system. In other words, it assumes a
\emph{quantum nondemolition procedure}. However, this assumption about premeasurement is
not necessarily true. In fact, as we will show, the pointer
states, which are characterized by their ability not to entangle with the states of another subsystem, generally are time-dependent. So, we must differentiate between the pointer state of a subsystem and the preferred basis of measurement. In this paper we refer to the
preferred basis of measurement as time-independent pointer states which can be realized only in certain regimes.

\subsection{Example: evolution of the two-level atom in the Jaynes-Cummings model}

To clarify this better here we present a
physical example, represented by the evolution of the two-level atom
in the Jaynes-Cummings model (JCM) of quantum optics, in order to show how
in practice a diagonal state with \emph{time-dependent} pointer states
can be created as a result of the natural evolution of the global
state of the system and the environment.

Consider the state of the two-level atom in the Jaynes-Cummings
model of quantum optics which involves a two-level atom, with upper
and lower levels that can respectively be represented by $
    |a\rangle
$ and $
    |b\rangle
$, interacting with a single-mode quantized electromagnetic field
inside an ideal cavity, represented by creation and annihilation
operators $\hat{a}^{\dag}$ and $\hat{a}$. For exact resonance and in
the rotating-wave approximation, the interaction Hamiltonian for the
composite system can be written as
\begin{equation}\label{6}
H_{I}=\hbar \rmg(\hat{a}^{\dag}\sigma_{-}+\sigma_{+}\hat{a}).
\end{equation}
Where
$\rmg=-\varrho_{12}.\hat{\epsilon}\sqrt{\frac{w}{2\hbar\varepsilon_{\circ}V}}$
 is the atom-field coupling constant, with $\varrho_{12}=e\langle
a|\textbf{r}|b\rangle $ as the atomic electric-dipole transition
matrix element. ($\hat{\epsilon}$ is the field polarization
vector, $\omega$ is the atomic transition frequency which is
taken to be resonant with the frequency of the cavity eigenmode, and V is the cavity
mode volume). Also $\sigma_{+}$ and $\sigma_{-}$ are the atomic flipping operators
given by
\begin{equation}\label{6}
\sigma_{+}=|a\rangle\langle b| \qquad \rmand \qquad
\sigma_{-}=|b\rangle\langle a|.
\end{equation}

Consider the field to be initially in the coherent state
$|\nu\rangle$
\begin{equation}\label{7}
    |\Phi_{\rm
    field}(t_{0})\rangle=|\nu\rangle=\sum_{n=0}^{\infty}\rmc_{n}|n\rangle;
    \quad\rmwith \quad
    \rmc_{n}=\frac{\rme^{-\frac{1}{2}|\nu|^{2}}\nu^{n}}{\sqrt{n!}},
\end{equation}
where $|\nu|^{2}=\bar{n}$ is the average number of photons in the
coherent state, and $\nu=|\nu|e^{-i\phi}$. In general, the exact
solution for an initial atomic state $|\psi_{\rm
atom}(t_{0})\rangle=\alpha|a\rangle+\beta|b\rangle$ and a field
state initially prepared in the coherent state, is a highly
entangled state of the field and the atom \cite{Scully}. However,
Gea-Banacloche \cite{Gea-Banacloche} has shown that for a large average number of
photons, if we consider the evolution of the initial atomic states
$|+\rangle$ and $|-\rangle$, defined by
\begin{equation}\label{8}
    |\pm\rangle=\frac{1}{\sqrt{2}}(\rme^{-\rmi\phi}|a\rangle\pm|b\rangle)
\end{equation}
(here $\phi$ is the same as the phase of $\nu=|\nu|e^{-i\phi}$), the evolution of the global
state of the system and the field (the environment) would be very
interesting. Gea-Banacloche proved that when the initial atom-field
state is $|\pm\rangle|\nu\rangle$, in the limit of
$\bar{n}\rightarrow\infty$ the global state of the two-level atom (2LA) and the field will evolve as follows:
\begin{equation}\label{9}
    |\pm\rangle|\nu\rangle |_{t=0}\rightarrow \frac{1}{\sqrt{2}}(\rme^{-\rmi\phi}
    \rme^{\mp \rmi \rmg t/(2\sqrt{\bar{n}})}|a\rangle\pm|b\rangle)\times|\Phi_{\pm}(t)\rangle,
\end{equation}
where
\begin{equation}\label{10}
    |\Phi_{\pm}(t)\rangle=\rme^{-\bar{n}/2}\sum_{n=0}^{\infty}
    \frac{\bar{n}^{n/2}}{\sqrt{n!}}\rme^{-\rmi n\phi}\rme^{\mp
    \rmi \rmg t\sqrt{n}}|n\rangle,
\end{equation}
gives us the time evolution of the state of the field. This result
holds for any time, provided that $t$ goes to infinity slowly enough
to have $t/\bar{n}\rightarrow0$. Since the time scale for the JCM
revivals is $t_{\rmR}=2\pi\sqrt{\bar{n}}/\rmg$ \cite{Scully},
$t_{\rmR}/\bar{n}\rightarrow0$ as $\bar{n}\rightarrow\infty$ (a
typical value for $\rmg/2\pi$ is 44 kHz in a micromaser experiment
\cite{Gea-Banacloche}). Hence, the approximate solution in equation (10) holds
accurately over a large number of revivals, as long as $\bar{n}$ is
large enough.

The states $|+\rangle$ and $|-\rangle$ form a basis set for the
two-level atom (2LA); therefore, the evolution of any other initial
atomic state with an initial coherent field can be expressed as a
linear combination of the evolution of $|+\rangle|\nu\rangle$ and
$|-\rangle|\nu\rangle$.
\begin{eqnarray}\label{11}
    \nonumber(\gamma|+\rangle+\delta|-\rangle)\ |\nu\rangle|_{t=0}\rightarrow
    \gamma\ |+(t)\rangle\ |\Phi_{+}(t)\rangle+\delta\ |-(t)\rangle\ |\Phi_{-}(t)\rangle
    \\ \rmwith  \qquad |+(t)\rangle=\frac{\rme^{-\rmi\phi}
    \rme^{-\rmi \rmg t/(2\sqrt{\bar{n}})}|a\rangle+|b\rangle}{\sqrt{2}} \\
    \rmand \nonumber \qquad |-(t)\rangle=\frac{\rme^{-\rmi\phi}\rme^{+\rmi
    \rmg t/(2\sqrt{\bar{n}})}|a\rangle-|b\rangle}{\sqrt{2}}.
\end{eqnarray}

The time-dependent states $|\pm(t)\rangle$ and $|\Phi_{\pm}(t)\rangle$ appearing in the above equations are \emph{the pointer states of the system (the 2LA) and the environment (the field) which are characterized by their ability not to entangle with each other}. As we observe,
in the limit of large $\bar{n}$ in which equation
(9) is valid, the global state of the 2LA and the
field remains as a product state if the atom is initially prepared
in one of the states $|\pm\rangle$. This means that, one can at all
times assign a well-defined pure state to the atom initially
prepared in one of the states $|\pm\rangle$ and clearly, no other
initial atomic states have this characteristic, as is obvious from
equation (11). In other words, for an initial coherent field and in
the limit of large $\bar{n}$, when the 2LA is initially prepared in
one of the two states $|\pm\rangle$, the field and the atom never
entangle; while equations (9) and (11) indicate that for any initial atomic
state other than the $|\pm\rangle$ states, the states of the field
and the atom will not remain separated and they entangle throughout
the interaction. Also, equation (11) indicates that for an arbitrary
initial atomic state and in the limit of large $\bar{n}$, there is always a one-to-one correspondence
between a (preferred) set of pointer states from the system (the 2LA) and some
corresponding states of a field which is initially prepared in a coherent state.

In essence, we observe that in
the limit of large $\bar{n}$ and for an initial coherent field, in
fact the coherent field does a Von Neumann premeasurement on the
state of the 2LA; which makes the global state in a
diagonal form and superselects a preferred set of
pointer states of the system. However, here the premeasurement
by the field definitely is not an ideal premeasurement, as the
initial atomic states $|\pm\rangle$ evolve by acquiring a
phase factor $\rme^{\mp\rmi \rmg t/(2\sqrt{\bar{n}})}$; except for
$t\ll t_{\rmR} $, for which this change is negligible and the right
hand side of equation (11) can be approximated by $\gamma\
|+\rangle|\Phi_{+}(t)\rangle+\delta\ |-\rangle|\Phi_{-}(t)\rangle$.

\subsection{Schmidt states versus pointer states}

It can be shown that in the limit of $\bar{n}\rightarrow\infty$, which
corresponds to the classical limit for which equations (9) to (11)
are valid, the field states $|\Phi_+ (t)\rangle$ and $|\Phi_-
(t)\rangle$ almost promptly become orthogonal \cite{Gea-Banacloche}. However, as is obvious from equation (11), the pointer states of the system are not orthogonal at all times; and hence the diagonal state of the total composite system (represented by equation (11)), which is created after premeasurement, cannot represent a Schmidt decomposition at all times; as by definition the Schmidt states of the system and the environment must be orthogonal amongst themselves.

The Schmidt states obtained by diagonalizing the density matrix of the system at each instant of time necessarily are not the same as the pointer states at corresponding times; even in certain regimes and those times long enough so that the pointer states of the system and the environment can be considered as orthogonal amongst themselves. (For our example of the JCM the pointer states of the system are almost orthogonal provided that $t$ goes to infinity slowly enough to have $t/\sqrt{\bar{n}}\rightarrow0$, as can be seen from equation (11); also the pointer states of the environment, given by equation (10), become orthogonal within a time of the order of $1/\rmg$ due to decoherence, as was proved by Gea-Banacloche \cite{Gea-Banacloche}).
This is basically because of the fact that the Schmidt basis obtained this way will not necessarily exhibit the quasiclassical properties which are characteristic of the dynamical pointer states of measurement. In fact, as we described, the pointer states of the system emerge dynamically as those states that do not entanglement with the environment; while the Schmidt states (which generally are not unique) necessarily are not robust against the entanglement with the environment. Hence, the pointer states of measurement generally cannot be obtained simply by diagonalizing the instantaneous density matrix of the system (however, it is shown that only when the Schmidt states of the system are very nearly degenerate they can be significantly different from those environment-selected pointer states which are orthogonal amongst themselves (i.e.\ pointer states at certain regimes and sufficiently long times, so that they can be considered as orthogonal); while they are almost the same as the environment-selected pointer states whenever they are far from degeneracy and the pointer states of the system and the environment can be considered as orthogonal amongst themselves. The interested reader for example can refer to the interesting article by Albrecht \cite{Albrecht}. This result in fact is just as we expect from the condition for the uniqueness of the Schmidt decomposition; since when the Schmidt states of the system are very nearly degenerate, this indicates that not all of the expansion coefficients in the diagonal state of the total composite system (equation (2)) are different and hence the Schmidt states are not unique. Therefore, in this case the Schmidt states which we obtain by some procedure necessarily will not be the same as the instantaneous pointer states of measurement which likewise diagonalize the state of the total composite system, but in addition to that do not entangle with the states of another subsystem at a subsequent time).

\subsection{The commutativity criterion}

In fact, the measurement cannot be considered faithful if the one-to-one correspondence between the states of the system and the apparatus is not preserved. In other words, if the
further interaction with an outer environment does not maintain the one-to-one
correspondence between the states of two subsystems, no longer can we have a faithful measurement. For example if
at an initial time $t_{0}$ the states of the system which appear in the diagonal state of the total composite system are the two states $|\psi_{1}(t_{0})\rangle$
and $|\psi_{2}(t_{0})\rangle$, then they can be considered as the instantaneous pointer states of the system if their further interaction with the environment preserves the one-to-one correspondence with the environment. In other words their evolution must be of the following form
\begin{eqnarray}\label{12}
    |\psi_{1}(t_{0})\rangle|E_{0}\rangle\rightarrow|\psi_{1}(t)\rangle|E_{1}(t)\rangle
    \quad \rmand \quad
    |\psi_{2}(t_{0})\rangle|E_{0}\rangle\rightarrow|\psi_{2}(t)\rangle|E_{2}(t)\rangle.
\end{eqnarray}
Now if we consider a superposition of these states at an initial time $t_{0}$ like
\begin{equation}\label{13}
|\psi_{\pm}(t_{0})\rangle=\alpha\ |\psi_{1}(t_{0})\rangle\pm\beta\ |\psi_{2}(t_{0})\rangle, \quad \rmwith \quad\alpha\beta\neq0
\end{equation}
then, due to the interaction with the environment such a state will
evolve according to
\begin{equation}\label{14}
|\psi_{\pm}(t_{0})\rangle|E_{0}\rangle\rightarrow\alpha\ |\psi_{1}(t)\rangle
|E_{1}(t)\rangle\pm\beta\ |\psi_{2}(t)\rangle|E_{2}(t)\rangle.
\end{equation}
This means that any superposition of $|\psi_{1}(t_{0})\rangle$ and
$|\psi_{2}(t_{0})\rangle$ states (the states which appear in the diagonal
state of the total composite system \emph{and} do not entangle with the
states of the environment) immediately entangles with the environment
and hence it will lose its individuality and become unobservable. Indeed, the pointer states
of the system emerge dynamically as those states that are the least
sensitive, or the most robust, to the interaction with the
environment; in the sense that they do not entangle with the
environment. This is commonly referred to as the \emph{stability
criterion} for the selection of the pointer states \cite{Schlosshauer1,Zurek1,Zurek2}. In essence, some states are robust in spite of the environmental interaction, while other states rapidly entangle with the environment, lose their individuality and therefore become unobservable in practice. However, the information about only those states of the system that do not entangle with the environment can be passed all the way to
the observer and these are the pointer states of measurement.

As we already mentioned by studying the evolution of the 2LA in the Jaynes-Cummings model of quantum optics, the pointer states characterized by their ability not to entangle with the environment, generally are \emph{time-dependent} and hence, in general we should distinguish between
the set of (time-dependent) pointer states and the preferred basis
of measurement. In section 3 we discuss why indeed the states which may be able to satisfy the requirement of faithful measurement generally are expected to be time-dependent. Also, in section 4 we will discuss the exact conditions under which the pointer states of measurement can be time-independent, so that a preferred set of basis states can exist as the basis of measurement.

With this introduction, now our ultimate goal is to find the preferred basis of measurement. However, we must first identify the pointer states of the system for an arbitrary total Hamiltonian defining the system-environment model. As we described, our general
selection criterion is given by the \emph{stability criterion}; i.e.\ the set of the pointer states of the system is given by those
states of the system that do not entangle with the environment. In other words, they keep their individuality; so that they are able to hold a one-to-one correspondence with the states of the environment. In
order to find these states, we should look for system states
$|s_{i}(t)\rangle$ (like those of equation (11) for the JCM) such that the composite system-environment state, when starting from a product state
$|s_{i}(t_{0})\rangle|E_{0}\rangle$ at $t=0$, remains in the product form
$|s_{i}(t)\rangle|E_{i}(t)\rangle$ at all subsequent times $t>0$ under
the action of the total Hamiltonian. Now we show that the commutativity
criterion for the determination of the pointer states of measurement, first introduced by Zurek \cite{Zurek1,Zurek2}, although is sufficient in order to assure the requirement of having a faithful measurement, it is too restrictive and generally does not hold in all situations. In other words, it is not the
minimal condition for having a faithful measurement.

Two regimes are often considered. In the \emph{quantum measurement limit} the interaction between
the system and the environment is so strong as to dominate the
evolution of the system. Therefore, in this limit it is assumed that the intrinsic
dynamics of the system and the environment is negligible in
comparison with the evolution induced by the interaction. i.e.
\begin{equation}\label{15}
    \hat{H}\approx \hat{H}_{\rm int};
\end{equation}
and hence, the evolution of the composite system-environment state is
approximately given by the evolution operator $\rme^{-i\int_{0}^{t}\hat{H}_{\rm int}(t')\ dt'}$.

The other limit which also sometimes is considered, corresponds to the case that the Hamiltonian for the system almost dominates the interaction between the system and the environment as well as the self-Hamiltonian of the environment. Hence, in this limit which frequently is called as the \emph{quantum limit of decoherence}, the following approximation is assumed
\begin{equation}\label{16}
    \hat{H}\approx \hat{H}_{\cal S}.
\end{equation}
As we show in section 4, the result of these approximations is that the pointer states of the system turn out to be independent of time (unlike those of the JCM in equation (11) which are obtained for the ``exact-resonance" regime).

In his famous 1981 paper Zurek argued that in the quantum measurement limit the preferred
set of pointer states for the system should be given by those states of the system
that are eigenstates of the part of the interaction Hamiltonian
$\hat{H}_{\rm int}$ pertaining to the Hilbert space of the system; since in this case we have
\begin{eqnarray}\label{17}
     \rme^{-\rmi\int_{0}^{t}\hat{H}_{\rm int}(t')\ dt'}|s_{i}\rangle|E_{0}\rangle=|s_{i}\rangle\
     \rme^{-\rmi\int_{0}^{t}(\lambda_{i}(t')\hat{E})\ dt'} |E_{0}\rangle\equiv
    |s_{i}\rangle|E_{i}(t)\rangle,
\end{eqnarray}
provided $\hat{H}_{\rm int}=\hat{S}\otimes\hat{E}$; with $\hat{S}$ and $\hat{E}$ denoting some operators in the Hilbert space of the system and the environment respectively; and $\hat{S}\ |s_{i}\rangle=\lambda_{i}|s_{i}\rangle$. As we see from equation (17), in this case the state of the system does not
entangle with the state of the environment. (If we consider the more general form of the interaction Hamiltonian given by $\hat{H}_{\rm int}=\sum_{\alpha} \hat{S}_{\alpha}\otimes\hat{\cal E}_{\alpha}$, then a sufficient condition for $\{|s_{i}\rangle\}$ to form a set of pointer states of the system is that the $|s_{i}\rangle$ be simultaneous eigenstates of \emph{all} the system operators $\hat{S}_{\alpha}$.)

Equivalently, if we define the pointer observable as the observable for the system whose
eigenstates are these pointer states $|s_{i}\rangle$ of the
system, i.e.\
\begin{equation}\label{18}
    \hat{O}_{\cal S}=\sum_{i}o_{i}|s_{i}\rangle\langle s_{i}|,
\end{equation}
since the $|s_{i}\rangle$ are eigenstates of $\hat{H}_{\rm int}$, it
follows that $\hat{O}_{\cal S}$ must commute with $\hat{H}_{\rm
int}$,
\begin{equation}\label{19}
    [\hat{O}_{\cal S},\hat{H}_{\rm int}]=0.
\end{equation}
This condition is often referred to as the \emph{commutativity
criterion} and was first discussed by Zurek in his paper of
1981 \cite{Zurek1}. However, this is not the only possible situation for identifying the pointer states of a system (i.e.\ as we will discuss here, in many situations the pointer states of the system, which are characterized by their ability not to entangle with the environment, are time-dependent and do not satisfy the commutativity criterion); since we note that \emph{if} the pointer states of the system are eigenstates of the Hamiltonian, as Zurek has proposed, for an interaction Hamiltonian which does not explicitly depend on time (like that of our example of the JCM at the zero-detuning regime) the most that can change about the pointer states of the system under the effect of the evolution operator is an overall phase factor and basically they would remain unaltered under the effect of the evolution operator. In other words, Zurek's pointer states are basically independent of time. However, the point is that in order to satisfy the requirement of faithful measurement the pointer states of the system do not have to be independent of time like Zurek's pointers. For example a one-to-one correspondence between some well-defined states of the system and some states from the environment is preserved in our example represented by the JCM (equation (11)); while as we saw, the pointer states which appear in the diagonal state of the total composite system in this example clearly depend on time.

Indeed, if we have pointer states which change by more than an overall phase factor (as those of the JCM), they cannot be eigenstates of the Hamiltonian at all times; and as one can easily verify, the pointer states of the system in the JCM, given by equation (11), also are not the eigenstates of the total Hamiltonian nor the interaction Hamiltonian at all times; although they satisfy the requirement of faithful measurement and do not entangle with the environment during the interaction. As a result, although the commutativity criterion fulfills the requirement of having a faithful measurement, we notice that it is not the minimal condition for determining the pointer states of measurement; in the sense that it is not always valid and we might have time-dependent pointer states like those of the JCM which do not satisfy any kind of commutation relation (including the total Hamiltonian of the composite system or the interaction Hamiltonian in the quantum measurement limit) at all times.

\subsection{Bloch vector and determination of the preferred basis of measurement}

The reduced density matrix of a two-level system $\hat{\rho}_{\cal S}(t)$ generally can be expressed in terms of the Bloch vector $\textbf{R}(t)\equiv(R_{x}, R_{y}, R_{z})$ \cite{Eberly} as follows
\begin{equation}\label{20}
\hat{\rho}_{\cal S}(t)=\frac{1}{2}(\hat{I}+\textbf{R}(t).\hat{\mathbf{\sigma}})=\frac{1}{2}(\hat{I}+R_{x}\sigma_{x}+R_{y}\sigma_{y}+R_{z}\sigma_{z});
\end{equation}
from which one can easily verify that the Bloch vector components must be defined by
\begin{equation}\label{21}
R_{x}=\rho_{ab}+\rho_{ba} \qquad R_{y}=i(\rho_{ab}-\rho_{ba})\quad \rmand \quad R_{z}=\rho_{aa}-\rho_{bb}.
\end{equation}
In the above equation we used the notation $\rho_{ab}=\langle a|\hat{\rho}_{\cal S}(t)|b\rangle$ (with $|a\rangle$ and $|b\rangle$ representing a complete set of basis states for the two-level system) and etc.

The Bloch vector here can be interpreted az the polarization of the state of the two-level system. This is because the direction of $\mathbf{R}$ tells us into what set of eigenstates the reduced density matrix of the system can be decomposed. For example, if $R_{x}=R_{y}=0$ and $R_{z}\neq0$ then
\begin{equation}\label{22}
\hat{\rho}_{\cal S}(t)=\frac{1}{2}(\hat{I}+R_{z}\sigma_{z})=\frac{1}{2}\left(
                                                                    \begin{array}{cc}
                                                                      1+R_{z} & 0 \\
                                                                      0 & 1-R_{z} \\
                                                                    \end{array}
                                                                  \right);
\end{equation}
therefore, $\hat{\rho}_{\cal S}$ will commute with $\sigma_{z}$ and can be decomposed in terms of the eigenstates of $\sigma_{z}$. As a result, generally speaking in a certain regime of the parameter space only if the components of the Bloch vector settle in some asymptotic values at $t\rightarrow\infty$, can we conclude that at sufficiently long times there can exist a preferred basis of measurement represented by the eigenstates of $\hat{\rho}_{\cal S}$. Otherwise, i.e.\ if the Bloch vector constantly changes its direction, a preferred basis of measurement cannot be realized in the corresponding regime.

As an example, Gea-Banacloche has studied the Bloch sphere evolution of the two-level atom in the Jaynes-Cummings model of quantum optics and for an initial coherent field \cite{Gea-Banacloche2}. As one can see from his studies, away from the specific features which he studied  (such as collapses and revivals and state preparation in the evolution of the two-level atom), the Bloch vector does not have any asymptotic behavior in most of the regimes. In paper IV \cite{paper4} we find those regimes of the parameter space for this model for which the Bloch vector has an asymptotic behavior for large times; so that a preferred basis of measurement can be realized.

As another example, consider the spin-spin model described by the total Hamiltonian
\begin{equation}\label{23}
\hat{H}=\hat{H}_{\cal S}+\hat{H}_{\rm int}=-\frac{1}{2}\Delta_{0}\hat{\sigma}_{x}+\frac{1}{2}\hat{\sigma}_{z}\otimes\sum_{i=1}^{N}\rmg_{i}\hat{\sigma}_{z}^{(i)},
\end{equation}
where the first term (representing the self-Hamiltonian of the central system) accounts for the intrinsic dynamics of the central spin-half particle and $\Delta_{0}$ is the so-called tunneling matrix element. Also, the second term in equation (23) corresponds to the linear interaction between the $\hat{\sigma}_{z}$ coordinate of the central spin and the $\hat{\sigma}_{z}$ coordinates of $N$ environmental spin-half particles, with coupling strengths $\rmg_{i}$.

The evolution of the Bloch vector has been studied for this model by Cucchietti et al. \cite{Cucchietti}. They considered a Gaussian spectral density for the initial state of the environmental spins and considered two main regimes. In the regime that the self-Hamiltonian of the system is negligible compared to the interaction between the system and the environment they found that $\mathbf{R}(t\rightarrow\infty)\rightarrow R_{z}$; while for the regime that the interaction between the system and the environment is negligible compared to the self-Hamiltonian of the system they found that $\mathbf{R}(t\rightarrow\infty)\rightarrow R_{x}$. This basically shows that in the first regime the preferred basis of measurement is determined as the eigenstates of the $\hat{\sigma}_{z}$ operator, while in the second regime they are determined as the eigenstates of the $\hat{\sigma}_{x}$ operator; just in agreement with the predictions of the commutativity criterion for the quantum measurement limit and the quantum limit of decoherence.

\subsection{Other methods for determination of the pointer states of measurement}

As we discussed, the pointer states of the system keep their individuality; as they do not entangle with the states of the environment and hold a one-to-one correspondence with the pointer states of the environment (see equation (12)). Therefore, if one initially prepares the system in one of its pointer states $|s_{i}\rangle$, on a further observation on the state of the environment he can expect only to observe the corresponding pointer state of the environment $|e_{i}\rangle$ (and vice versa). In this sense pointer states are \emph{predictable}; just as classical states are.

Also we discussed that the commutativity criterion for determination of the pointer states cannot be a reliable criterion in many of the realistic cases; because pointer states, characterized by their ability not to entangle with the states of another subsystem, generally are time-dependent (we will mathematical prove this latter point within the next section). Realizing the unreliability of the commutativity criterion for determination of the pointer states of measurement, some researchers have tried to introduce a better predictability criterion \cite{Zurek2005,Zurek93a,ZHP}. However, these other criteria for determination of the pointer states of measurement more or less suffer from the same limitations of the commutativity criterion; again because pointer states generally are time-dependent. Hence, we may have the problem of stability of solutions with respect to time. Moreover, the mathematical calculations required while using these methods often are quite difficult; even for some very simple models.

Among these methods, Zurek has introduced the so-called ``predictability sieve" \cite{Zurek2005,Zurek93a,ZHP}, which exploits the Von Neumann entropy to measure the loss of predictability caused by evolution. In this case pointer states correspond to the least entropy producing states and predictability is a function of time and a functional of the initial state of the system. So, pointer state are sought by maximizing the predictability functional over the initial state of the system. Also, the purity of a system $\rm Tr[\rho^{2}](t)$ (more exactly ``purity loss time") has been exploited to measure the loss of predictability in the so-called ``purity sieve" \cite{Zurek2005}.

One main issue regarding the different predictability sieves which have been introduced so far is that there is no \emph{a priori} reason to expect that all of these criteria lead to the same set of pointer states; although in the macroscopic limit the difference between various sieves are expected to be negligible \cite{117}. However, the authors believe that the problem of the stability of solutions for pointer states (which arises from the fact that the pointer states of measurement generally are time-dependent) and the calculational difficulties involved in these methods are the main issues with these methods.

\subsection{Another aspect of pointer states: redundant encoding of information in the environment}

In some of the more recent publications the main focus is on the information encoded in the environment from the state of the system and the spread of this information through the environment; rather than the system environment interactions. In this approach the environment acquires the role of a communication channel and the pointer states of the system correspond to those states of the system whose imprints on the subsystems of the environment are \emph{most redundantly} and \emph{most robustly} encoded in the environment. In other words, the main focus of this kind of approach is the transformation of the information encoded in an ensemble of environmental ``witness states" all the way to the observer. This process indeed is another step of the measurement process. However, we do not focus on this aspect of the measurement process in this research.

The research on this aspect of pointer states is carried out under the headings of ``the environment as a witness" and ``quantum Darwinism"; and it has been claimed by Ollivier \cite{110,111} and also by Blume-Kohout and Zurek \cite{112,113} that the environment-selected pointer states of the system not only are the states which are the least entangled with the states of the environment (i.e.\ the states which are the most robust against the environmental interactions) but also they are the states which can be imprinted most completely and most redundantly in many distinct subsets of the environment.\\

\section{Identifying time-dependent Pointer States of measurement for an arbitrary Hamiltonian}

In order to be able to obtain the pointer states of the system and the environment for an arbitrary total Hamiltonian defining the system-environment model we first need to find those probable initial states of the system which do not entangle with the states of the environment throughout their evolution with time; and then we should obtain their time evolution. Finally, we should obtain their corresponding states from the environment which in fact, are the pointer states of the environment. As we will see in this section, existence of pointer states often requires having a sufficiently large environment which contains a large number of degrees of freedom. In other words, pointer states characterized by their ability not to entangle with the states of another subsystem, do not necessarily exist in any arbitrary regime.

After developing our method in this section we will exploit it in order to rederive the time-dependent pointer states of the two-level-atom and the field (initially prepared in the coherent state) in the JCM and for the exact resonance regime. As we will see, the previous results obtained by Gea-Banacloche (equations (8) to (11)) are easily obtained using our method. Also, in paper $\sf III$ \cite{paper3} and paper $\sf IV$ \cite{paper4} we will show how easily we can use this method in order to obtain the time-dependent pointer states of the system and the environment for the generalized spin-boson model (SBM) and also for the JCM and in some nonresonance regimes.


Consider a two-state system $\cal S$ with two arbitrary basis states $|a\rangle$ and $|b\rangle$, initially prepared in the state
\begin{equation}\label{24}
|\psi^{\cal S}(t_{0})\rangle=\alpha|a\rangle+\beta|b\rangle \quad \rmwith \quad |\alpha|^{2}+|\beta|^{2}=1,
\end{equation}
and an environment initially prepared in the state
\begin{equation}\label{25}
|\phi^{\cal E}(t_{0})\rangle=\sum_{n=0}^{\infty}c_{n}|\varphi_{n}\rangle,
\end{equation}
where $\{|\varphi_{n}\rangle\}$'s are a complete set of basis states for the environment.
For the two-state system with the two basis states $|a\rangle$ and $|b\rangle$ we can take the set of any four linearly independent operators in the Hilbert space of the system as a complete set of basis operators, which can induce any change to the initial state of the two-state system given by equation (24). For example, we can take the Pauli operators in addition to the identity operator $\hat{I}=|a\rangle\langle a|+|b\rangle\langle b|$ as our complete set of basis operators; or equivalently we can take the four operators $|a\rangle\langle a|$, $|a\rangle\langle b|$, $|b\rangle\langle a|$ and $|b\rangle\langle b|$ as our complete set of basis operators. So, the time evolution operator for the global state of the system and the environment, which (for a two-state system) generally is of the form
\begin{equation}\label{26}
\hat{U}_{\rm tot}(t)=\sum_{\alpha=1}^{4} \hat{S}_{\alpha}\otimes\hat{\cal E}_{\alpha}\ ,
\end{equation}
Can be considered as
\begin{equation}\label{27}
\hat{U}_{\rm tot}(t)=\hat{\cal E}_{1}|a\rangle\langle a|+\hat{\cal E}_{2}|a\rangle\langle b|
+\hat{\cal E}_{3}|b\rangle\langle a|+\hat{\cal E}_{4}|b\rangle\langle b|.
\end{equation}
In the above equation $\hat{\cal E}_{i}$'s depend on the total Hamiltonian defining the system-environment model. For example, for the Jaynes-Cummings model of quantum optics and for exact resonance and in the rotating wave approximation (RWA), it can be shown \cite{Scully} that the $\hat{\cal E}_{i}$'s are given by the following relations
\begin{eqnarray}\label{28}
\nonumber \hat{\cal E}_{1}=\cos(\rmg t\sqrt{\hat{a}^{\dag}\hat{a}+1})\ , \quad \hat{\cal E}_{2}=-i\ \frac{\sin(\rmg t\sqrt{\hat{a}^{\dag}\hat{a}+1}}{\sqrt{\hat{a}^{\dag}\hat{a}+1}}\ \hat{a})\\
\hat{\cal E}_{3}=-i\hat{a}^{\dag}\ \frac{\sin(\rmg t\sqrt{\hat{a}^{\dag}\hat{a}+1}}{\sqrt{\hat{a}^{\dag}\hat{a}+1}})\ , \quad
\hat{\cal E}_{4}=\cos(\rmg t\sqrt{\hat{a}^{\dag}\hat{a}}).
\end{eqnarray}

Using equations (24) to (27) we can write the global state of the system and the environment as follows
\begin{eqnarray}\label{29}
\nonumber |\psi^{\rm tot}(t)\rangle=\hat{U}_{\rm tot}(t).\ (\alpha|a\rangle+\beta|b\rangle)\otimes(\sum_{n=0}^{\infty}c_{n}|\varphi_{n}\rangle)\\
=\textbf{A}(t)\ |a\rangle+\textbf{B}(t)\ |b\rangle\ \quad \rmwith \quad
\textbf{A}(t)=\sum_{n=0}^{\infty}c_{n}\{\alpha\hat{\cal E}_{1}+\beta\hat{\cal E}_{2}\}\ |\varphi_{n}\rangle \\ \rmand \quad \nonumber
\textbf{B}(t)=\sum_{n=0}^{\infty}c_{n}\{\alpha\hat{\cal E}_{3}+\beta\hat{\cal E}_{4}\}\ |\varphi_{n}\rangle.
\end{eqnarray}
In order to find those probable initial states of the system which do not entangle with the states of the environment we first define $\hat{G}(t)$ as the operator in the Hilbert space of the environment which relates the vectors $\textbf{A}(t)$ and $\textbf{B}(t)$ to each other
\begin{equation}\label{30}
\textbf{A}(t)=\hat{G}(t)\textbf{B}(t) \ \ \rmor \quad \sum_{n}c_{n}\{\alpha\hat{\cal E}_{1}+\beta\hat{\cal E}_{2}\}\ |\varphi_{n}\rangle=\hat{G}(t){\sum_{n}c_{n}\{\alpha\hat{\cal E}_{3}+\beta\hat{\cal E}_{4}\}\ |\varphi_{n}\rangle}.
\end{equation}
Now, for the global state of the system and the environment, which is given by
\begin{eqnarray}\label{31}
\nonumber |\psi^{\rm tot}(t)\rangle=\textbf{A}(t)\ |a\rangle+\textbf{B}(t)\ |b\rangle=\hat{G}(t)\textbf{B}(t)\ |a\rangle+\textbf{B}(t)\ |b\rangle\\=\{\hat{G}(t)|a\rangle+|b\rangle\}\times(\sum_{n=0}^{\infty}c_{n}\{\alpha\hat{\cal E}_{3}+\beta\hat{\cal E}_{4}\}\ |\varphi_{n}\rangle),
\end{eqnarray}
we observe that \emph{if} for some initial states of the system and the environment $\hat{G}(t)$ turns out to become in the form
\begin{eqnarray}\label{30.1}
\hat{G}(t)=G(t)\times\hat{I}_{\cal E},
\end{eqnarray}
with $G(t)$ as a scalar (rather than an operator) and $\hat{I}_{\cal E}$ representing the identity operator in the Hilbert space of the environment, then those initial states of the system and the environment will not entangle with each other, and hence they can represent the initial pointer states of the system and the environment. This result simply is because of the fact that if for some initial states of the system and the environment $\hat{G}(t)$ turns out to become a scalar in the form of equation (\ref{30.1}), $G(t)$ will be independent of the indices of the environment (i.e.\ independent of $n$); as in this case all components of $\textbf{B}(t)$ will be mapped into their corresponding components from $\textbf{A}(t)$ through the \emph{same} scalar function $G(t)$ (which will keep the two vectors $\textbf{A}(t)$ and $\textbf{B}(t)$ parallel to each other). Therefore, in this case $\hat{G}(t)$ will not enter the summation in the expression $\sum_{n}c_{n}\{\alpha\hat{\cal E}_{3}+\beta\hat{\cal E}_{4}\}\ |\varphi_{n}\rangle$ of equation (\ref{31}); and (as one can see from equation (\ref{31})) the states of the system and the environment respectively represented by $\{G(t)|a\rangle+|b\rangle\}$ and $\sum_{n}c_{n}\{\alpha\hat{\cal E}_{3}+\beta\hat{\cal E}_{4}\}\ |\varphi_{n}\rangle$ will not entangle to each other. In other words, if for some initial states of the system and the environment the operator $\hat{G}(t)$ becomes proportional to the identity operator, the two vectors $\textbf{A}(t)$ and $\textbf{B}(t)$ will stay parallel with each other throughout their evolution with time, and the states of the system and the environment will not entangle with each other; therefore (as one can see from equation (\ref{31})), in this case pointer states can be realized for the system and the environment given by
\begin{eqnarray}\label{32}
\nonumber |\pm(t)\rangle=\calN\ \{G(t)|a\rangle+|b\rangle\} \qquad \rmand \\
|\Phi_{\pm}(t)\rangle=\calN^{-1}(\sum_{n=0}^{\infty}c_{n}\{\alpha\hat{\cal E}_{3}+\beta\hat{\cal E}_{4}\}\ |\varphi_{n}\rangle).
\end{eqnarray}
In the above equation we have represented the pointer states of the system by $|\pm(t)\rangle$ and those of the environment by $|\Phi_{\pm}(t)\rangle$. Also, $\cal N$ is the normalization factor for the pointer states of the system (clearly $\cal N=\rm\frac{1}{\sqrt{2}}$ if $|G(t)|=1$, as we will see for the example of JCM in exact resonance and the rotating wave approximation).

As we will see in this paper, generally there is no guaranty for the condition (\ref{30.1}) to be satisfied; and satisfaction of this condition often may require having a sufficiently large environment which contains a large number of
degrees of freedom. However, \emph{if} in some regime and for a given Hamiltonian defining a system-environment model we can find initial states for the system and the environment which satisfy this condition, we do know that pointer states can be realized for the system and the environment and these initial states would correspond to the initial pointer states of the system and the environment.

In order to find the pointer states of the system and the environment for a given total Hamiltonian defining
the system-environment model, and for a given initial state of the environment, our main goal would be finding those possible initial states of the system for which $\hat{G}(t)$ (which is defined through equation (\ref{30})) is of the form of relation (\ref{30.1}). Within the following paragraphs we will consider our previous example of the JCM with an initial coherent field and exploit this method in order to rederive the time-dependent pointer states of the two-level-atom and the field, which we already saw in equations (10) and (11). As we will see, this task is not as difficult as it might initially seem and finding initial states of the system which make the operator $\hat{G}(t)$ proportional to the identity operator in the Hilbert space of the environment often can be done quite easily when dealing with a sufficiently large environment which contains a large number of degrees of freedom. However, for an initial state of the environment which does not correspond to a sufficiently large environment (as we will see) we might not have any initial states which can satisfy our condition (\ref{30.1}) for determining the pointer states of the system and the environment. This means that pointer states, which are characterized by their ability not to entangle with the states  of another subsystem, do not necessarily exist in any arbitrary regime.

For our example of the JCM if we use the number states as the complete set of basis states for the environment, then equation (28) for the environmental operators ($\hat{\cal E}_{i}$'s) would suggest us that in this case we have
\begin{eqnarray}\label{33}
\nonumber \hat{\cal E}_{1}|\varphi_{n}\rangle=f_{1}(n,t)|\varphi_{n}\rangle, \ \ \ \ \ \ \qquad \hat{\cal E}_{2}|\varphi_{n}\rangle=f_{2}(n,t)|\varphi_{n-1}\rangle \\
\hat{\cal E}_{3}|\varphi_{n}\rangle=f_{3}(n,t)|\varphi_{n+1}\rangle \quad \rmand \quad \hat{\cal E}_{4}|\varphi_{n}\rangle=f_{4}(n,t)|\varphi_{n}\rangle;
\end{eqnarray}
with
\begin{eqnarray}\label{34}
\nonumber f_{1}(n,t)= \cos(\rmg t\sqrt{n+1}),\ \ \ \ \qquad f_{2}(n,t)=-i\sin(\rmg t\sqrt{n}) \\
f_{3}(n,t)=-i\sin(\rmg t\sqrt{n+1}) \quad \rmand \quad f_{4}(n,t)=\cos(\rmg t\sqrt{n}).
\end{eqnarray}
Using equation (\ref{33}) and our definition of the operator $\hat{G}(t)$ (equation (\ref{30})), we can write
\begin{equation}\label{35}
\sum_{n}\{\alpha c_{n}f_{1}(n,t)+\beta c_{n+1}f_{2}(n+1,t)\}\ |\varphi_{n}\rangle=\hat{G}(t)\sum_{n}\{\alpha c_{n-1}f_{3}(n-1,t)+\beta c_{n}f_{4}(n,t)\}\ |\varphi_{n}\rangle.
\end{equation}
The above relation is valid whenever equation (\ref{33}) is valid for a total Hamiltonian defining the system-environment model (as for our example of the JCM).

Now, for the pointer states $\hat{G}(t)$ must satisfy the condition (\ref{30.1}) for obtaining the pointer states of the system and the environment, i.e.\ $\hat{G}(t)=G(t)\times\hat{I}_{\cal E}$. Therefore, since $\{|\varphi_{n}\rangle\}$ is a complete set of basis states for the environment, for the initial pointer states we can open the summations in equation (\ref{35}) and equalize terms from the two sides of this equation which correspond to the same basis state $|\varphi_{n}\rangle$ to obtain
\begin{equation}\label{36}
G(t)=\frac{\{\alpha c_{n}f_{1}(n,t)+\beta c_{n+1}f_{2}(n+1,t)\}}{\{\alpha c_{n-1}f_{3}(n-1,t)+\beta c_{n}f_{4}(n,t)\}}; \quad \forall\ n>0.
\end{equation}

The above result for $G(t)$ which generally depends on $n$ would contradict our initial assumption of $\hat{G}(t)$ being a scalar \emph{unless} if we can find certain initial states for the system for which $G(t)$ turns out to become independent of $n$. \footnote{We would like to see if the condition can be satisfied for \emph{any} initial state of the system and the environment with $G(t)$ becoming independent of the index $n$ of the states of the environment. So, if finally we can find any specific set of initial states for the system and the environment which satisfies this condition with $G(t)$ independent of the indices of the environment, then we have reached our goal and our assumption has not been in vain.} So now we should seek for those particular initial states of the system which can make $G(t)$ independent of the index $n$ of the states of the environment. For this purpose we assume the field to be initially in the coherent state and use the $f_{i}(n,t)$ functions from equation (\ref{34}) in order to simplify equation (\ref{36}) for our example of the JCM in exact resonance and the RWA:
\begin{equation}\label{37}
G(t)=\frac{\{\alpha c_{n}\cos(\rmg t\sqrt{n+1})-i\beta c_{n}e^{-i\varphi}\sqrt{\frac{\bar{n}}{n+1}}\sin(\rmg t\sqrt{n+1})\}}{\{\beta c_{n}\cos(\rmg t\sqrt{n})-i\alpha c_{n}e^{i\varphi}\sqrt{\frac{n}{\bar{n}}}\sin(\rmg t\sqrt{n})\}};
\end{equation}
as for the coherent field (represented by equation (7)) we have $c_{n+1}=c_{n}e^{-i\varphi}\sqrt{\frac{\bar{n}}{n+1}}$ and $c_{n-1}=c_{n}e^{i\varphi}\sqrt{\frac{n}{\bar{n}}}$\ . However, in the limit of a large average number of photons $\bar{n}\rightarrow\infty$ we can replace the factors $\sqrt{\frac{n}{\bar{n}}}$ and $\sqrt{\frac{\bar{n}}{n+1}}$ by unity,
since the Poisson distribution of the coherent field is extremely sharp for $\bar{n}\rightarrow\infty$ and hence, (in the summations of equation (\ref{35})) for $\bar{n}\rightarrow\infty$ and $n\approx\bar{n}$ we have $\sqrt{\frac{n}{\bar{n}}}\approx1$ and $\sqrt{\frac{\bar{n}}{n+1}}\approx1$, while for $n$ being far from $\bar{n}$ the corresponding $c_{n}$ coefficients are negligible. As a result, equation (\ref{37}) for $G(t)$ can be further simplified to
\begin{equation}\label{38}
G(t)=\frac{\{\alpha c_{n}\cos(\rmg t\sqrt{n+1})-i\beta c_{n}e^{-i\varphi}\sin(\rmg t\sqrt{n+1})\}}{\{\beta c_{n}\cos(\rmg t\sqrt{n})-i\alpha c_{n}e^{i\varphi}\sin(\rmg t\sqrt{n})\}}.
\end{equation}
In fact, as Gea-Banacloche has shown \cite{Gea-Banacloche}, the difference between such approximate expressions which are obtained by assuming $\sqrt{\frac{n}{\bar{n}}}\approx1$ and $\sqrt{\frac{\bar{n}}{n+1}}\approx1$ and the exact expressions (where we keep these factors) goes to zero as $\bar{n}\rightarrow\infty$.
Moreover, in this limit we can use
\begin{equation}\label{39}
\sqrt{n+1}-\sqrt{n}\approx\frac{1}{2\sqrt{\bar{n}}}
\end{equation}
as one can easily verify for example by writing the Taylor expansion of $\sqrt{n}$ about $\bar{n}$.

Now, from equation (\ref{38}) clearly we have $\alpha=\beta G(0)$; and hence by this substitution we find
\begin{equation}\label{40}
G(t)=\frac{G(0)\cos(\rmg t\sqrt{n+1})-i\ e^{-i\varphi}\sin(\rmg t\sqrt{n+1})}{\cos(\rmg t\sqrt{n})-i\ G(0)\ e^{i\varphi}\sin(\rmg t\sqrt{n})}.
\end{equation}
By looking at the above equation one would easily see that if $G(0)=\pm e^{-i\varphi}$ (i.e.\ if $\alpha=\pm \beta e^{-i\varphi}$), $G(t)$ will be independent of the index $n$ of the states of the environment; since in this case we have
\begin{equation}\label{41}
G(t)=\pm e^{-i\varphi}\ \frac{e^{\mp i\rmg t\sqrt{n+1}}}{e^{\mp i\rmg t\sqrt{n}}}=\pm e^{-i\varphi}\ e^{\mp i\rmg t(\sqrt{n+1}-\sqrt{n})}\ ;
\end{equation}
however, at the limit of a large average number of photons we can use equation (\ref{39}) to replace the factor $\sqrt{n+1}-\sqrt{n}$ by $\frac{1}{2\sqrt{\bar{n}}}$ and find
\begin{equation}\label{42}
G(t)=\pm e^{-i(\varphi\pm\rmg t/2\sqrt{\bar{n}})}
\end{equation}
which clearly is independent of the index $n$ of the states of the environment.

This result simply means that for $\alpha=\pm \beta e^{-i\varphi}$ which is equivalent to having
\begin{equation}\label{43}
\alpha=\frac{e^{-i\varphi}}{\sqrt{2}} \quad \rmand \quad \beta=\pm\frac{1}{\sqrt{2}}
\end{equation}
(since we must have $|\alpha|^{2}+|\beta|^{2}=1$) the states of the system and the environment will not entangle with each other. Moreover, using equation (\ref{32}) which gives us the general time evolution of the pointer states of the system; and $G(t)$ of equation (\ref{42}) (which is independent of the index $n$ of the states of the environment) we find the time evolution of the pointer states of the system as follows
\begin{equation}\label{44}
|\pm(t)\rangle=\frac{e^{- i\varphi}
    e^{\mp i \rmg t/(2\sqrt{\bar{n}})}|a\rangle\pm|b\rangle}{\sqrt{2}}.
\end{equation}

Next, let us use equation (\ref{32}) in order to obtain the corresponding pointer states of the environment; i.e.\ $|\phi_{\pm}(t)\rangle$. Here, substituting $\alpha$ and $\beta$ from equation (\ref{43}) and using equations (\ref{33}) and (\ref{34}) we have
\begin{eqnarray}\label{45}
\nonumber |\phi_{\pm}(t)\rangle=\sqrt{2}\ \sum_{n=0}^{\infty}c_{n}\{\alpha\hat{\cal E}_{3}+\beta\hat{\cal E}_{4}\}\ |\varphi_{n}\rangle \\
=\sum_{n=0}^{\infty}c_{n}\{\mp i\ e^{- i\varphi}\sin(\rmg t\sqrt{n+1})\ |n+1\rangle+\cos(\rmg t\sqrt{n})\ |n\rangle\}.
\end{eqnarray}
(In writing the above equation for mathematical convenience we made use of the fact that an overall phase is not important in determining the pointer states of the environment and only the relative phases are important). However, for the coherent field we had $c_{n+1}=c_{n}e^{-i\varphi}\sqrt{\frac{\bar{n}}{n+1}}$; also, as we already discussed, at the limit of $\bar{n}\rightarrow\infty$ we can neglect the factor $\sqrt{\frac{\bar{n}}{n+1}}$ and replace $c_{n}\rme^{-\rmi\varphi}$ of the above equation by $c_{n+1}$ to have
\begin{eqnarray}\label{46}
\nonumber |\phi_{\pm}(t)\rangle=\sum_{n=0}^{\infty}\{\mp i\ c_{n+1}\ \sin(\rmg t\sqrt{n+1})\ |n+1\rangle+c_{n}\cos(\rmg t\sqrt{n})\ |n\rangle\} \\
=\sum_{n=0}^{\infty}c_{n}\{\mp i\ \sin(\rmg t\sqrt{n})+\cos(\rmg t\sqrt{n})\}|n\rangle=\sum_{n=0}^{\infty}c_{n}e^{\mp i\rmg t\sqrt{n}}\ |n\rangle,
\end{eqnarray}
where in the second line of the above equation we used the substitution $n\longrightarrow n-1$ and made use of the fact that the first term in the first summation is equal to zero and hence we can keep the lower limit of the summation as $n=0$.

This way we easily reproduced the previous results first introduced by Gea-Banacloche (given by equations (10) and (11)) through using our general method for finding the pointer states of the system and the environment. We will further demonstrate the generality and usefulness of this method in our other papers where we obtain the time-dependent pointer states of the system and the environment
for the generalized spin-boson model \cite{paper3} and also for the JCM and in some nonresonance regimes \cite{paper4} (these are new results not containing in previous works). However, the significance of this formulation is not only because of providing us with a method for obtaining the pointer states of the system and the environment for a given total Hamiltonian. In fact, as we discuss in more detail in the next section, this formulation specifically is useful because of the insight which it can bring us regarding the general properties of pointer states under different regimes and circumstances; and more importantly the insight which it brings us regarding the question of determination of the preferred basis of measurement.

\section {Determination of the preferred basis of measurement}

We already showed that \emph{if} in some regime and for some specific values of $\alpha$ and $\beta$ of the initial state of the system the function $\hat{G}(t)$, defined by equation (\ref{30}), turns out independent of the states of the environment then in that regime the corresponding values for $\alpha$ and $\beta$ are related to the initial pointer states of the system, which will not entangle with the states of the environment. Moreover, the time evolution of the pointer states of the system and the environment can be obtained with the help of these initial values of $\alpha$ and $\beta$ and by using equation (\ref{32}). Nevertheless, by looking at equation (\ref{30}) which defines $G(t)$ we notice that as the evolution operators $\hat{\cal E}_{i}$ (which correspond to the Hilbert space of the environment) generally are time-dependent, generally we expect the function $\hat{G}(t)$ and therefore the pointer states of the system (given by equation (\ref{32})) to be dependent on time.

The above consideration basically means that the states from the system which might be able not to entangle with the states of another subsystem necessarily are not independent of time and hence we must search for the special conditions under which we might have time-independent pointer states, so that a preferred set of basis states can be identified as the basis of measurement. However, before discussing our criteria for identifying time-independent pointer states and their further consequences here we briefly discuss an instructive physical example where the pointer states of the system turn out to be time-independent.

Our example is represented by the simplified spin-boson model (SBM) which is composed of a central spin-half particle surrounded by an environment of $N$ bosonic particles. For our simplified spin-boson model we consider a single-mode quantized field for the environment; and moreover, we disregard a possible contribution to the self-Hamiltonian of the system which can induce transitions between the upper and lower states of the central system (i.e.\ an intrinsic tunneling contribution proportional to $\hat{\sigma}_{x}$ Pauli matrix that would generate the intrinsic dynamics of the central spin). So, we consider the following total Hamiltonian for our model
\begin{equation}\label{47}
\hat{H}=\frac{1}{2}\omega_{0}\hat{\sigma}_{z}+\omega\hat{a}^{\dag}\hat{a}+\hat{\sigma}_{z}\otimes(\rmg \hat{a}^{\dag}+\rmg^{\ast}\hat{a});
\end{equation}
where in the above equation $\omega_{0}$ is the splitting between the states of the spin-half particle and $\omega$ is the frequency of the cavity eigenmode. The third term, with $\rmg$ as the spin-field coupling constant, represents the interaction between the central spin-half particle and a single-mode quantized field; which in fact is the quantized form of the famous
$-\vec{\mu}.\textbf{B}$
Hamiltonian due to the interaction between a particle of magnetic dipole-moment $\vec{\mu}$ and a magnetic field $\textbf{B}$.

This model has been studied by many people (for an interesting review the reader can refer to the article by Leggett et al. \cite{Leggett} or Schlosshauer's book \cite{Schlosshauer1}) and as one can easily show, the effective evolution operator in the interaction-picture for this model can be represented by
\begin{eqnarray}\label{48}
\nonumber \hat{V}(t)=\rmexp\{\hat{\sigma}_{z}\otimes(\lambda(t)\ \hat{a}^{\dag}-\lambda^{\ast}(t)\ \hat{a})\}\\ \rmwith \quad \lambda(t)=\frac{\rmg}{\omega}(1-e^{i\omega t})
\end{eqnarray}
(the different notations for the time evolution operator between this example and our general formulation is just because our effective evolution operator, given by equation (\ref{48}), in fact differs from the actual evolution operator by an overall phase; which of course is not physically important). The key point regarding this evolution operator is that it contains \emph{only one} of the Pauli spin operators (here $\hat{\sigma}_{z}$; as a result of which  $\hat{\sigma}_{z}$ becomes a constant of motion in this simplified model). As we will see in the following paragraphs, this will result in having \emph{time-independent} pointer states given by the eigenstates of $\hat{\sigma}_{z}$\ .

Now if we consider the initial state of the total composite system as
\begin{equation}\label{49}
|\Psi^{\rmtot}(t_{0})\rangle=(\alpha|a\rangle+\beta|b\rangle)\ |\Phi^{\cal E}\rangle,
\end{equation}
with $|a\rangle$ and $|b\rangle$ representing the eigenstates of $\hat{\sigma}_{z}$ and $|\Phi^{\cal E}\rangle$ representing some arbitrary initial state of the environment, then using the evolution operator given by equation (\ref{48}) we easily obtain
\begin{eqnarray}\label{50}
\nonumber |\Psi^{\rmtot}(t)\rangle=\hat{V}(t)\ |\Psi^{\rmtot}(t_{0})\rangle=\alpha|a\rangle|\Phi_{+}(t)\rangle+\beta|b\rangle|\Phi_{-}(t)\rangle
\\ \nonumber \rmwith \quad |\Phi_{+}(t)\rangle=\hat{D}(\lambda(t))\ |\Phi^{\cal E}\rangle  \\ \rmand \quad\ |\Phi_{-}(t)\rangle=\hat{D}(-\lambda(t))\ |\Phi^{\cal E}\rangle.
\end{eqnarray}
In the above equation $\hat{D}(\lambda(t))$, which generates the evolution of the pointer states of the environment, is defined by
\begin{equation}\label{51}
\hat{D}(\lambda(t))=\rmexp[\lambda(t)\ \hat{a}^{\dag}-\lambda^{\ast}(t)\ \hat{a}]
\end{equation}
(which in fact is the same as the displacement operator in quantum optics). The concrete form of the environmental pointer states $|\Phi_{\pm}(t)\rangle$ clearly would depend on our initial state of the environment $|\Phi^{\cal E}\rangle$. However, no matter what is our initial state of the environment here we clearly observe that the interaction between the system and the environment would select the time-independent eigenstates of $\hat{\sigma}_{z}$ as pointer states of the system, which will be robust against the entanglement with the environment.

We could equivalently arrive at this result by using our method for obtaining the pointer states of the system and the environment. (The following discussion might seem excessive. However, we are going through it in order to get into the roots of having time-independent pointer states and then finally relate the insight which we obtain through these examples to our formulation for calculating the pointer states of measurement; and find out the possible conditions under which the pointer states of the system turn out to become independent of time; so that they can represent the preferred basis of measurement). In fact, we can expand the exponential in equation (\ref{48}) to write
\begin{eqnarray}\label{52}
\nonumber \hat{V}(t)=\rmexp\{\hat{\sigma}_{z}\otimes(\lambda(t)\ \hat{a}^{\dag}-\lambda^{\ast}(t)\ \hat{a})\}\\=\sum_{n=0}^{\infty}\frac{1}{n!}\ (\hat{\Lambda}(t))^{n}\ \hat{\sigma}_{z}^{n},
\end{eqnarray}
where in the second line of the above equation we defined $\hat{\Lambda}(t)=\lambda(t)\ \hat{a}^{\dag}-\lambda^{\ast}(t)\ \hat{a}$. But we have $\hat{\sigma}_{z}^{2l}=\hat{I}$ and $\hat{\sigma}_{z}^{2l+1}=\hat{\sigma}_{z}$. So, now we can write
\begin{eqnarray}\label{53}
\nonumber \hat{V}(t)=\sum_{l=0}^{\infty}\frac{1}{(2l)!}\ (\hat{\Lambda}(t))^{2l}\ \hat{I}+\sum_{l=0}^{\infty}\frac{1}{(2l+1)!}\ (\hat{\Lambda}(t))^{(2l+1)}\ \hat{\sigma}_{z} \\
=\cosh(\hat{\Lambda}(t))\ \hat{I}+\sinh(\hat{\Lambda}(t))\ \hat{\sigma}_{z}.
\end{eqnarray}
However, the last expression can be simplified as
\begin{eqnarray}\label{54}
\nonumber \hat{V}(t)=\exp(\hat{\Lambda}(t))|a\rangle\langle a|+\exp(-\hat{\Lambda}(t))|b\rangle\langle b| \\=\hat{D}(\lambda(t))\ |a\rangle\langle a|+\hat{D}(-\lambda(t))\ |b\rangle\langle b|
\end{eqnarray}
Comparing this result with our general form for the evolution operator, given by equation (\ref{27}), we find
\begin{equation}\label{55}
 \hat{\cal E}_{1}=\hat{D}(\lambda(t))\ , \quad \hat{\cal E}_{2}=\hat{\cal E}_{3}=0 \quad \rmand \quad \hat{\cal E}_{4}=\hat{D}(-\lambda(t)).
\end{equation}
 Here we show that having time-\emph{independent} pointer states given by the states $|a\rangle$ and $|b\rangle$ (i.e.\ the eigenstates of the $\hat{\sigma}_{z}$ operator, which are the basis states that we used in order to write our evolution operator in the form of equation (27)) is the result of having $\hat{\cal E}_{2}=\hat{\cal E}_{3}=0$ in this example; and this is a general condition. i.e.\

\noindent \textbf{\emph{Theorem 1}}: \emph{Whenever} in some basis $\hat{\cal E}_{2}=\hat{\cal E}_{3}=0$, then those basis states will be time-independent pointer states of the system.

\noindent \emph{\textbf{Proof:}} If $\hat{\cal E}_{2}=\hat{\cal E}_{3}=0$,
operating the evolution operator $\hat{U}_{\rmtot}(t)=\hat{\cal E}_{1}|a\rangle\langle a|+\hat{\cal E}_{4}|b\rangle\langle b|$ on the initial state of the total composite system $|\psi^{\rmtot}(t_{0})\rangle=(\alpha|a\rangle+\beta|b\rangle)\otimes|\Phi^{\cal E}(t_{0})\rangle$ we obtain
\begin{equation}\label{57}
|\psi^{\rmtot}(t)\rangle=\alpha|a\rangle\ \hat{\cal E}_{1}\ |\Phi^{\cal E}(t_{0})\rangle+\beta|b\rangle\ \hat{\cal E}_{4}\ |\Phi^{\cal E}(t_{0})\rangle.
\end{equation}
The above relation basically means if the system initially is prepared in the $|a\rangle$ state (i.e.\ $\beta=0$), the evolution of the total composite system must be given by $|\psi^{\rmtot}(t_{0})\rangle=|a\rangle\otimes|\Phi^{\cal E}(t_{0})\rangle\rightarrow|a\rangle\ \hat{\cal E}_{1}\ |\Phi^{\cal E}(t_{0})\rangle$ and if the system initially is prepared in the $|b\rangle$ state (i.e.\ $\alpha=0$), the evolution of the total composite system must be given by $|\psi^{\rmtot}(t_{0})\rangle=|b\rangle\otimes|\Phi^{\cal E}(t_{0})\rangle\rightarrow|b\rangle\ \hat{\cal E}_{4}\ |\Phi^{\cal E}(t_{0})\rangle$. In other words, the basis states $|a\rangle$ and $|b\rangle$ are the time-independent pointer states of the system; as they do not entangle with the states of the environment.

However, in this case the corresponding pointer states of the environment necessarily are not time-independent and are given by
\begin{equation}\label{62}
|\phi_{a}(t)\rangle=\hat{\cal E}_{1}\ |\Phi^{\cal E}(t_{0})\rangle \quad \rmand \quad |\phi_{b}(t)\rangle=\hat{\cal E}_{4}\ |\Phi^{\cal E}(t_{0})\rangle.
\end{equation}
QED.

Here we note that although requiring the condition $\hat{\cal E}_{2}=\hat{\cal E}_{3}=0$ would guarantee having time-independent pointer states in the interaction picture given by our initial basis states, these states can represent the preferred basis of measurement only if either $\hat{H}_{\cal S}\approx0$ or they turn out to be eigenstates of the self-Hamiltonian of the system as well; since in general an arbitrary state of a system $|\alpha\rangle$ in the interaction picture is related to that of the Schr\"{o}dinger picture by $|\alpha;t\rangle_{\rm S}=e^{-i\hat{H}_{0}t}\ |\alpha;t\rangle_{\rm I}$. For our example of the simplified spin-boson model this further condition is satisfied and hence the basis states $|a\rangle$ and $|b\rangle$ do represent the preferred basis of measurement in this case.

We continue through studying the conditions under which the pointer states of the system may become independent from time by presenting three more theorems. In each theorem we present a condition for having time-independent pointer states and predict the corresponding stationary pointer states. We will relate the mathematical conditions of theorem 1 and theorem 2 for having time-independent pointer states to the symmetries in the Hamiltonian of the total composite system through theorem 3 and theorem 4; which will be our main physical criteria for predicting the preferred basis of measurement. We will discuss the significance of these new results more elaborately in our final conclusion (section 5), where we better clarify how these results can serve as a generalization of the existing theory for determination of the preferred basis of measurement.

\noindent \textbf{\emph{Theorem 2}}:
Whenever in some basis $|a\rangle$ and $|b\rangle$ for the state of a two-level system we have $\hat{\cal E}_{1}=\hat{\cal E}_{4}$ and $\hat{\cal E}_{2}=e^{-i\varphi}\hat{\cal E}_{3}$, then we will have a pair of time-independent pointer states for the system given by $|\pm\rangle=\frac{1}{\sqrt{2}}\{|a\rangle\pm e^{i\frac{\varphi}{2}}|b\rangle\}$. e.g. if $\varphi=0$ ($\hat{\cal E}_{2}=\hat{\cal E}_{3}$) and the basis states $|a\rangle$ and $|b\rangle$ represent the eigenstates of the $\hat{\sigma}_{z}$ operator, then our time-independent pointer states in the interaction picture should be represented by the eigenstates of the $\hat{\sigma}_{x}$ operator.

\noindent \textbf{\emph{Proof}}: If $\hat{\cal E}_{1}=\hat{\cal E}_{4}$ and $\hat{\cal E}_{2}=e^{-i\varphi}\hat{\cal E}_{3}$, then the condition for obtaining the pointer states of the system and the environment reads
\begin{eqnarray}\label{63}
\nonumber \sum_{n}c_{n}\{\alpha\hat{\cal E}_{1}+\beta\hat{\cal E}_{2}\}\ |\varphi_{n}\rangle=\hat{G}(t)\sum_{n}c_{n}\{\alpha\ e^{i\varphi}\hat{\cal E}_{2}+\beta\hat{\cal E}_{1}\}\ |\varphi_{n}\rangle \quad \rmand \\ \hat{G}(t)\ be\ proportional\ to\ the\ unit\ matrix.
\end{eqnarray}
The above condition can be satisfied for $\alpha=\pm e^{-i\frac{\varphi}{2}}\beta$; since it is obvious from equation (\ref{63}) that for $\alpha=\pm e^{-i\frac{\varphi}{2}}\beta$ we have $\hat{G}(t)=\pm e^{-i\frac{\varphi}{2}}\times\hat{I}_{\cal E}$. So, from equation (\ref{32}) we see that in this case the pointer states of the system must be given by
\begin{equation}\label{64}
|\pm(t)\rangle=\calN\  \{G(t)|a\rangle+|b\rangle\}=\frac{1}{\sqrt{2}}\{|a\rangle\pm e^{i\frac{\varphi}{2}}|b\rangle\}
\end{equation}
QED.

In paper $\sf IV$ \cite{paper4} we will show that the above condition with $\varphi=0$ (i.e.\ $\hat{\cal E}_{1}=\hat{\cal E}_{4}$ and $\hat{\cal E}_{2}=\hat{\cal E}_{3}$) is satisfied while studying the evolution of the two-level atom in the Jaynes-Cummings model of quantum optics and in the regime that $\hat{H}_{\cal S}\approx0$ and $\hat{H}_{\cal E}\ll\hat{H}_{\rm int}$ (but $\hat{H}_{\cal E}$ necessarily is not equal to zero). As a result, in this regime of $\hat{H}_{\cal S}\approx0$ our theorem predicts that the preferred basis of measurement must be given by the eigenstates of the $\hat{\sigma}_{x}$ operator. Interestingly, for this example and in this regime $\hat{H}_{\rm int}$ turns out to be proportional to $\hat{\sigma}_{x}$ and hence the Zurek theorem for determination of the preferred basis of measurement \emph{also} predicts the preferred basis of measurement to be given by the eigenstates of $\hat{\sigma}_{x}$ (which are the eigenstates of the total Hamiltonian in the quantum measurement limit $\hat{H}\approx \hat{H}_{\rm int}$). In fact, as we will show, the above theorem always covers the predictions of Zurek's theory for determination of the preferred basis of measurement at corresponding limits; although, as we will see, it is much more general compared to the former theory.

Now, a very interesting question can be to ask: ``How can we predict whether any of the conditions given by the above two theorems can be satisfied for an arbitrary total Hamiltonian defining a system-environment model?" More specifically, ``Can we predict any of these conditions from the symmetries in the Hamiltonian of the total composite system"? Being able to answer the above question is specifically important, since it is not always easy to calculate the time evolution operator for a given total Hamiltonian defining a system-environment model. So, it can be quite useful if we can obtain some information about the preferred basis of measurement before calculating the evolution operator and knowing the $\hat{\cal E}_{i}$ operators. In fact, if we can answer the above question for an arbitrary total Hamiltonian defining a system-environment model, we will be able to predict those regimes of the parameter space (the parameter space for example determines how big is the contribution of each term in the Hamiltonian of the total composite system) for which the pointer states of the system can become independent from time so that a preferred basis of measurement can be realized. Moreover, we will be able to predict the corresponding preferred basis of measurement for each regime.

In what follows we relate the mathematical conditions discussed in our aforementioned theorems for having time-independent pointer states to the symmetries in the Hamiltonian of the total composite system. We do this through the following two theorems:

\noindent \textbf{\emph{Theorem 3}}: In the total Hamiltonian of the global composite system $\hat{H}_{\rmtot}=\hat{H}_{\cal S}+\hat{H}_{\cal E}+\hat{H}^{'}$ if the interaction Hamiltonian between the system and the environment commutes with the self-Hamiltonian of the system, i.e.\ if $[\hat{H}_{\cal S},\hat{H}^{'}]=0$, we must have time-independent pointer states for the system given by the eigenstates of $\hat{H}_{\cal S}$ provided $\hat{H}_{\cal S}\neq0$ and $\hat{H}_{\cal S}\neq\hat{I}$; or the eigenstates of $\hat{H}^{'}$ if $\hat{H}_{\cal S}=0$ or $\hat{H}_{\cal S}=\hat{I}$.

\noindent \textbf{\emph{Proof}}: The Hamiltonian, in the interaction picture, is given by
\begin{equation}\label{65}
\hat{H}_{\rm int}=e^{i\hat{H}_{0}t}\hat{H}^{'}e^{-i\hat{H}_{0}t} \quad \rmwith \quad \hat{H}_{0}=\hat{H}_{\cal S}+\hat{H}_{\cal E}.
\end{equation}
So, using the Baker-Hausdorff Lemma i.e.\
\begin{equation}\label{66}
e^{\alpha \hat{A}}\hat{B}e^{-\alpha \hat{A}}=\hat{B}+\alpha[\hat{A},\hat{B}]+\frac{\alpha^{2}}{2!}[\hat{A},[\hat{A},\hat{B}]]+...
\end{equation}
we can clearly see that if $[\hat{H}_{\cal S},\hat{H}^{'}]=0$, then we must have $[\hat{H}_{\rm int},\hat{H}_{\cal S}]=0$. Therefore, the evolution operator in the interaction picture, which generally is given by
\begin{equation}\label{67}
\hat{U}(t)=\cal T_{\leftarrow}\ \rm e^{-i\int_{0}^{t}\hat{H}_{int}(t')\ dt'}
\end{equation}
(with $\cal T_{\leftarrow}$ representing the time-ordering operator in the above equation), also must commute with $\hat{H}_{\cal S}$. i.e.\
\begin{equation}\label{68}
[\hat{U}(t),\hat{H}_{\cal S}]=0.
\end{equation}
The above result simply means that if $[\hat{H}_{\cal S},\hat{H}^{'}]=0$ the
eigenstates of $\hat{H}_{\cal S}$ must be time-independent pointer states of the system (provided $\hat{H}_{\cal S}\neq\hat{I}$ and $\hat{H}_{\cal S}\neq0$); as they will be eigenstates of the evolution operator $\hat{U}(t)$ as well and cannot be changed by more than an overall phase factor under its effect\footnote{For example if $\hat{H}_{\cal S}$ and $\hat{H}^{'}$ are proportional to one of the Pauli Matrices $\hat{\sigma}_{i}$ (as in our example of the simplified SBM), then in principle $\hat{U}(t)$ can be expanded in terms of powers of $\hat{\sigma}_{i}$ (see equations (\ref{67}) and (\ref{48})) . However, $\hat{\sigma}_{i}^{\ 2l}$ (with $l$ representing an integer) is equal to $\hat{I}$ the identity operator and $\hat{\sigma}_{i}^{\ 2l+1}=\hat{\sigma}_{i}$ . This means that here in principle the evolution operator $\hat{U}(t)$ can be written as a summation with terms which either contain $\hat{\sigma}_{i}$ or the identity operator. Therefore, it must commute with $\hat{\sigma}_{i}$ and $\hat{H}_{\cal S}$; resulting in having the eigenstates of $\hat{H}_{\cal S}$ as time-independent pointer states of the system.}.

Also for the case that $\hat{H}_{\cal S}=\hat{I}$ or $\hat{H}_{\cal S}=0$ from equation (\ref{65}) we can see that the operator $\hat{H}_{\rm int}$ cannot change the eigenstates of $\hat{H}^{'}$ in the Hilbert space of the system by more than an overall phase factor. Therefore, the evolution operator $\hat{U}(t)$, given by equation (\ref{67}), also cannot change the eigenstates of $\hat{H}^{'}$ by more than an overall phase factor and the eigenstates of $\hat{H}^{'}$ in the Hilbert space of the system will be the time-independent pointer states of the system QED.

In what follows, we study the relationship between the condition $[\hat{H}_{\cal S},\hat{H}^{'}]=0$ and the mathematical conditions for having time-independent pointer states presented in our first two theorems.
In other words we establish the connection between theorem 3 and the first two theorems.

We already showed that for $[\hat{H}_{\cal S},\hat{H}^{'}]=0$ we must have $[\hat{U}(t),\hat{H}_{\cal S}]=0$. Now, let us study what are the implications of the latter commutation relation regarding the environmental operators $\hat{\cal E}_{i}$ which appear in the evolution operator of the global composite system represented by equation (\ref{27}).

The self-Hamiltonian of the two-level system, which here commutes with $\hat{H}^{'}$ (the interaction between the system and the environment), generally can be represented as
\begin{equation}\label{69}
\hat{H}_{\cal S}=s_{1}|a\rangle\langle a|+s_{2}|a\rangle\langle b|+s_{3}|b\rangle\langle a|+s_{4}|b\rangle\langle b|.
\end{equation}
In the above equation like always $|a\rangle$ and $|b\rangle$ are some basis states for the two-level system and $s_{i}$'s are some numbers (rather than operators). As an example, for the simplified spin boson model (SBM) $\hat{H}_{\cal S}$ and $\hat{H}^{'}$ both are taken to be proportional to the $\hat{\sigma}_{z}$ operator; so in this case $s_{1}=-s_{4}=1$ and $s_{2}=s_{3}=0$. Now, using equations (\ref{27}) and (\ref{69}) we can rewrite $[\hat{U}(t),\hat{H}_{\cal S}]=0$ as
\begin{eqnarray}\label{70}
\nonumber [\hat{U}(t),\hat{H}_{\cal S}]=[\hat{\cal E}_{1}\frac{(1+\hat{\sigma}_{z})}{2}+\hat{\cal E}_{4}\frac{(1-\hat{\sigma}_{z})}{2}+\hat{\cal E}_{2}\hat{\sigma}_{+}+\hat{\cal E}_{3}\hat{\sigma}_{-}\ ,\\ \qquad \qquad \ \ s_{1}\frac{(1+\hat{\sigma}_{z})}{2}+s_{4}\frac{(1-\hat{\sigma}_{z})}{2}+s_{2}\hat{\sigma}_{+}+s_{3}\hat{\sigma}_{-}]=0.
\end{eqnarray}
Simplifying the above $2\times2$ matrix relation we find an equivalent set of three equations given by
\begin{eqnarray}\label{71}
\nonumber s_{3}(\hat{\cal E}_{4}-\hat{\cal E}_{1})+\hat{\cal E}_{3}\ (s_{1}-s_{4})=0 \quad \rmand \\
s_{2}(\hat{\cal E}_{1}-\hat{\cal E}_{4})-\hat{\cal E}_{2}\ (s_{1}-s_{4})=0 \quad \rmand \\
\nonumber s_{3}\hat{\cal E}_{2}=s_{2}\hat{\cal E}_{3}.
\end{eqnarray}
In calculating equation (\ref{71}) from equation (\ref{70}) we used the following commutation relations
\begin{equation}\label{72}
[\hat{\sigma}_{+}, \hat{\sigma}_{-}]=\hat{\sigma}_{z} \quad \rmand \quad [\hat{\sigma}_{z}, \hat{\sigma}_{+}]=2\hat{\sigma}_{+} \quad \rmand \quad [\hat{\sigma}_{z}, \hat{\sigma}_{-}]=-2\hat{\sigma}_{-}\ .
\end{equation}
However, note that for $\hat{H}_{\cal S}$ to be Hermitian we must have $s_{2}=s_{3}^{\ast}$. So, if we represent the phase of $s_{2}$ by $\varphi/2$; i.e.\ if $s_{2}=|s_{2}|e^{-i\varphi/2}$, then $s_{3}\hat{\cal E}_{2}=s_{2}\hat{\cal E}_{3}$ would mean that
\begin{equation}\label{73}
\hat{\cal E}_{2}=\hat{\cal E}_{3}e^{-i\varphi}.
\end{equation}
So, equations (\ref{71}) would require $\hat{\cal E}_{2}=\hat{\cal E}_{3}e^{-i\varphi}$ (provided $s_{2}=s_{3}^{\ast}\neq0$); as well as either of the following situations
\begin{eqnarray}\label{74}
\nonumber \hat{\cal E}_{2}=\hat{\cal E}_{3}=0 \quad \rmand \quad \hat{\cal E}_{1}=\hat{\cal E}_{4} \quad \rmor \\
\nonumber \hat{\cal E}_{2}=\hat{\cal E}_{3}=0 \quad \rmand \quad s_{2}=s_{3}=0 \quad \rmor \\
\hat{\cal E}_{1}=\hat{\cal E}_{4} \quad \rmand \quad s_{1}=s_{4} \quad \rmor \quad \rm etc.
\end{eqnarray}
Now,

$(a)$ For the first one of the above conditions ($\hat{\cal E}_{2}=\hat{\cal E}_{3}=0 \ \rmand \ \hat{\cal E}_{1}=\hat{\cal E}_{4}$) we have $\hat{U}(t)=\hat{\cal E}_{1}(|a\rangle\langle a|+|b\rangle\langle b|)=\hat{\cal E}_{1}\hat{I}$. So, all possible states of the system including the eigenstates of $\hat{H}_{\cal S}$ (as well as the states $|a\rangle$ and $|b\rangle$ themselves, also due to theorem 1 as here $\hat{\cal E}_{2}=\hat{\cal E}_{3}=0$), will be time-independent in the interaction picture. However, as we pointed out, only those of the states which also are eigenstates of the self-Hamiltonian of the system can represent the preferred basis of measurement; since these are the only states which will not have any time-evolution in the Schr\"{o}dinger picture. This is just in agreement with what is stated by theorem 3, which predicts the preferred basis of measurement as the eigenstates of $\hat{H}_{\cal S}$ for the case that $[\hat{H}_{\cal S},\hat{H}^{'}]=0$.

$(b)$ The second possible condition resulting from $[\hat{H}_{\cal S},\hat{H}^{'}]=0$, given in equation (\ref{74}) by $\hat{\cal E}_{2}=\hat{\cal E}_{3}=0 \ \rmand \ s_{2}=s_{3}=0 $, is a special case of theorem 1 and hence it predicts having time-independent pointer states given by the eigenstates of $\hat{\sigma}_{z}$. However, here we have $s_{2}=s_{3}=0 $; so $\hat{H}_{\cal S}$ must be given by $\hat{H}_{\cal S}=s_{1}|a\rangle\langle a|+s_{4}|b\rangle\langle b|$; whose eigenstates are the basis states $|a\rangle$ and $|b\rangle$. So as we observe, the time-independent pointer states of the system predicted by theorem 1 for this case are the same as the eigenstates of the self-Hamiltonian of the system $\hat{H}_{\cal S}$; just in agreement with what is stated by theorem 3.

$(c)$ The third condition ($\hat{\cal E}_{1}=\hat{\cal E}_{4} \ \rmand \ s_{1}=s_{4}$) together with $\hat{\cal E}_{2}=\hat{\cal E}_{3}e^{-i\varphi}$ of equation (\ref{73}) (provided $s_{2}=s_{3}^{\ast}\neq0$) is a special case of theorem 2 and hence it predicts having time-independent pointer states for the system given by $|\pm\rangle=\frac{1}{\sqrt{2}}\{|a\rangle\pm e^{i\frac{\varphi}{2}}|b\rangle\}$. Moreover, $s_{1}=s_{4}$; so from equation (\ref{69}) it is clear that $\hat{H}_{\cal S}$ must be given by
$\hat{H}_{\cal S}=s_{1}\hat{I}+s_{2}\hat{\sigma}_{+}+s_{3}\hat{\sigma}_{-}$. Also, $s_{2}=s_{3}^{\ast}$ (and $s_{2}=|s_{2}|e^{-i\varphi/2}$); so we must have $s_{3}=s_{2}e^{i\varphi}$. Therefore,
\begin{equation}\label{75}
\hat{H}_{\cal S}=s_{1}\hat{I}+s_{2}\hat{\sigma}_{+}+s_{2}e^{i\varphi}\hat{\sigma}_{-}=\left(
                                                                               \begin{array}{cc}
                                                                                 s_{1} & s_{2} \\
                                                                                 s_{2}e^{i\varphi} & s_{1} \\
                                                                               \end{array}
                                                                             \right).
\end{equation}
It is easy to verify that the eigenstates of the above matrix are the same as $|\pm\rangle=\frac{1}{\sqrt{2}}\{|a\rangle\pm e^{i\frac{\varphi}{2}}|b\rangle\}$ which were predicted by theorem 2 to be the time-independent pointer states of the system for this case. So as we observe, for the third possible condition resulting from $[\hat{H}_{\cal S},\hat{H}^{'}]=0$ also theorem 2 predicts having time-independent pointer states for the system given by the eigenstates of the self-Hamiltonian of the system; just in agreement with what is stated by theorem 3.

As an example of having the condition discussed in theorem 3 suppose that both contributions $\hat{H}^{'}$ and $\hat{H}_{\cal S}$ are proportional to the $\hat{\sigma}_{z}$ operator (i.e.\ $s_{1}=-s_{4}=1$ and $s_{2}=s_{3}=0$). Then according to equation (\ref{71}) we must have $\hat{\cal E}_{2}=\hat{\cal E}_{3}=0$ and hence, as a result of theorem 1 we must have time-independent pointer states for the system given by the eigenstates of $\hat{\sigma}_{z}$. Similarly, if both Hamiltonians $\hat{H}^{'}$ and $\hat{H}_{\cal S}$ are proportional to $\hat{\sigma}_{x(y)}$, we find $\hat{\cal E}_{1}=\hat{\cal E}_{4}$ and $\hat{\cal E}_{2}=\pm\hat{\cal E}_{3}$ and hence, as a result of theorem 2 we must have time-independent pointer states for the system given by the eigenstates of $\hat{\sigma}_{x(y)}$.

Next, we discuss another possible symmetry in the Hamiltonian of a given system-environment model which also would lead to having time-\emph{independent} pointer states for the system.

\noindent \textbf{\emph{Theorem 4}}: For the Hamiltonian in the interaction picture, which in some basis $|a\rangle$ and $|b\rangle$ of the two-level system can be represented by
\begin{equation}\label{76}
\hat{H}_{\rm int}=\hat{h}_{11}|a\rangle\langle a|+\hat{h}_{12}|a\rangle\langle b|+\hat{h}_{21}|b\rangle\langle a|+\hat{h}_{22}|b\rangle\langle b|,
\end{equation}
we must have $\hat{\cal E}_{2}=\hat{\cal E}_{3} \ \rmand \ \hat{\cal E}_{1}=\hat{\cal E}_{4}$ and hence, (according to theorem 2) time-independent pointer states for the system given by $|\pm\rangle=\frac{1}{\sqrt{2}}\{|a\rangle\pm|b\rangle\}$ provided
\begin{equation}\label{77}
\hat{h}_{11}=\hat{h}_{22} \quad \rmand \quad \hat{h}_{12}=\hat{h}_{21}.
\end{equation}

\noindent \textbf{\emph{Proof}}: For the time evolution operator in the interaction picture, which satisfies the Schr\"{o}dinger equation, we have
\begin{eqnarray}\label{78}
\nonumber i\hbar\frac{\partial}{\partial t}\hat{u}(t)=\hat{H}_{\rm int}\hat{u}(t) \quad \rm i.e.\ \\
i\hbar \left(
         \begin{array}{cc}
           \dot{\hat{\cal E}_{1}} & \dot{\hat{\cal E}_{2}} \\
           \dot{\hat{\cal E}_{3}} & \dot{\hat{\cal E}_{4}} \\
         \end{array}
       \right)=\left(
                 \begin{array}{cc}
                   \hat{h}_{11} & \hat{h}_{12} \\
                   \hat{h}_{21} & \hat{h}_{22} \\
                 \end{array}
               \right)
       \left(
                 \begin{array}{cc}
                   \hat{\cal E}_{1} & \hat{\cal E}_{2} \\
                   \hat{\cal E}_{3} & \hat{\cal E}_{4} \\
                 \end{array}
               \right).
\end{eqnarray}
So, if at some regime of the parameter space we have $\hat{h}_{11}=\hat{h}_{22} \ \rmand \ \hat{h}_{12}=\hat{h}_{21}$ we would have the following set of four equations
\begin{eqnarray}\label{79}
\nonumber i\hbar\dot{\hat{\cal E}_{1}}=\hat{h}_{11}\hat{\cal E}_{1}+\hat{h}_{12}\hat{\cal E}_{3},\\
\nonumber i\hbar\dot{\hat{\cal E}_{2}}=\hat{h}_{11}\hat{\cal E}_{2}+\hat{h}_{12}\hat{\cal E}_{4},\\
i\hbar\dot{\hat{\cal E}_{3}}=\hat{h}_{11}\hat{\cal E}_{3}+\hat{h}_{12}\hat{\cal E}_{1},\\
\nonumber i\hbar\dot{\hat{\cal E}_{4}}=\hat{h}_{11}\hat{\cal E}_{4}+\hat{h}_{12}\hat{\cal E}_{2}.
\end{eqnarray}
Now, this set of four equations is invariant under the transformation $\hat{\cal E}_{1}\leftrightarrow\hat{\cal E}_{4}$ and $\hat{\cal E}_{3}\leftrightarrow\hat{\cal E}_{2}$. Moreover, $\hat{\cal E}_{1}$ and $\hat{\cal E}_{4}$ satisfy the same initial conditions; just as $\hat{\cal E}_{2}$ and $\hat{\cal E}_{3}$ do (since we must have $\hat{u}(t_{0})=\hat{I}$; therefore, $\hat{\cal E}_{1}(t_{0})=\hat{\cal E}_{4}(t_{0})=1$ and $\hat{\cal E}_{2}(t_{0})=\hat{\cal E}_{3}(t_{0})=0$). As a result, whatever are the solutions of the set of equations (\ref{79}) we must have $\hat{\cal E}_{2}=\hat{\cal E}_{3} \ \rmand \ \hat{\cal E}_{1}=\hat{\cal E}_{4}$. Therefore, (according to theorem 2) we predict having time-independent pointer states for the system given by $|\pm\rangle=\frac{1}{\sqrt{2}}\{|a\rangle\pm|b\rangle\}$ QED.

Here also, we note that although the condition represented by equation (\ref{77}) would guarantee having time-independent pointer states in the interaction picture given by $|\pm\rangle=\frac{1}{\sqrt{2}}\{|a\rangle\pm|b\rangle\}$, these states can represent the preferred basis of measurement only if either $\hat{H}_{\cal S}\approx0$ or they turn out to be eigenstates of the self-Hamiltonian of the system as well; since in general an arbitrary state of a system $|\alpha\rangle$ in the interaction pictures is related to that of the Schr\"{o}dinger picture by $|\alpha;t\rangle_{\rm S}=e^{-i\hat{H}_{0}t}\ |\alpha;t\rangle_{\rm I}$.

As we will show in another article, the condition in theorem 4 is satisfied for the Jaynes-Cummings model of quantum optics and in the regime that $\hat{H}_{\cal S}\approx0$ and $\hat{H}_{\cal E}\ll\hat{H}_{\rm int}$ (but $\hat{H}_{\cal E}$ necessarily is not equal to zero). In fact, an exact calculation of the pointer states of the two-level system in this regime and by using our method, would result in finding the states $|\pm\rangle=\frac{1}{\sqrt{2}}\{|a\rangle\pm|b\rangle\}$ (with $|a\rangle$ and $|b\rangle$ representing the atomic upper and lower states respectively) as the time-independent pointer states of the system; just as is predicted by theorem 4.

We should also mention that the above theorem can be generalized as follows:

\noindent \textbf{\emph{Generalization of  Theorem 4}}:  For the Hamiltonian in the interaction picture, which in some basis $|a\rangle$ and $|b\rangle$ of the two-level system can be represented by equation (\ref{76}) we must have $\hat{\cal E}_{2}=\hat{\cal E}_{3}e^{-i\varphi} \ \rmand \ \hat{\cal E}_{1}=\hat{\cal E}_{4}$ and hence, (according to theorem 2) time-independent pointer states for the system given by $|\pm\rangle=\frac{1}{\sqrt{2}}\{|a\rangle\pm e^{-i\varphi}|b\rangle\}$ provided
\begin{equation}\label{80}
\hat{h}_{11}=\hat{h}_{22} \quad \rmand \quad \hat{h}_{12}=\hat{h}_{21}e^{-i\varphi}.
\end{equation}

\noindent \textbf{\emph{Proof}}: Using equation (\ref{78}) for the case that $\hat{h}_{11}=\hat{h}_{22} \ \rmand \ \hat{h}_{12}=\hat{h}_{21}e^{-i\varphi}$ and then taking the second derivative with respect to time of the operators $\hat{\cal E}_{i}$ one can easily verify the following set of four equations
\begin{eqnarray}\label{81}
\nonumber i\hbar\ddot{\hat{\cal E}_{1}}=\{\dot{\hat{h}}_{11}+\frac{\hat{h}_{11}^{2}}{i\hbar}+\frac{\hat{h}_{12}^{2}}{i\hbar}e^{i\varphi}\}\hat{\cal E}_{1}+\{\dot{\hat{h}}_{12}+\frac{2\hat{h}_{11}\hat{h}_{12}}{i\hbar}\}\hat{\cal E}_{3},\\
\nonumber i\hbar\ddot{\hat{\cal E}_{2}}=\{\dot{\hat{h}}_{11}+\frac{\hat{h}_{11}^{2}}{i\hbar}+\frac{\hat{h}_{12}^{2}}{i\hbar}e^{i\varphi}\}\hat{\cal E}_{2}+\{\dot{\hat{h}}_{12}+\frac{2\hat{h}_{11}\hat{h}_{12}}{i\hbar}\}\hat{\cal E}_{4},\\
i\hbar\ddot{\hat{\cal E}_{3}}=\{\dot{\hat{h}}_{11}+\frac{\hat{h}_{11}^{2}}{i\hbar}+\frac{\hat{h}_{12}^{2}}{i\hbar}e^{i\varphi}\}\hat{\cal E}_{3}+\{\dot{\hat{h}}_{12}e^{i\varphi}+\frac{2\hat{h}_{11}\hat{h}_{12}}{i\hbar}e^{i\varphi}\}\hat{\cal E}_{1},\\
\nonumber i\hbar\ddot{\hat{\cal E}_{4}}=\{\dot{\hat{h}}_{11}+\frac{\hat{h}_{11}^{2}}{i\hbar}+\frac{\hat{h}_{12}^{2}}{i\hbar}e^{i\varphi}\}\hat{\cal E}_{4}+\{\dot{\hat{h}}_{12}e^{i\varphi}+\frac{2\hat{h}_{11}\hat{h}_{12}}{i\hbar}e^{i\varphi}\}\hat{\cal E}_{2}.
\end{eqnarray}
this set of four equations is invariant under the transformation $\hat{\cal E}_{1}\leftrightarrow\hat{\cal E}_{4}$ and $\hat{\cal E}_{3}\leftrightarrow\hat{\cal E}_{2}e^{i\varphi}$. Moreover, $\hat{\cal E}_{1}$ and $\hat{\cal E}_{4}$ satisfy the same initial conditions; just as $\hat{\cal E}_{2}$ and $\hat{\cal E}_{3}e^{-i\varphi}$ do (since we must have $\hat{u}(t_{0})=\hat{I}$; therefore, $\hat{\cal E}_{1}(t_{0})=\hat{\cal E}_{4}(t_{0})=1$ and $\hat{\cal E}_{2}(t_{0})=\hat{\cal E}_{3}(t_{0})=0$. Also, the initial time $t_{0}$ is the same for all $\hat{\cal E}_{i}$'s; so in solving the set of equations (\ref{81}) we should not worry about the equality of the other constant of integration). As a result, whatever are the solutions of the set of equations (\ref{81}) we must have $\hat{\cal E}_{2}=\hat{\cal E}_{3}e^{-i\varphi} \ \rmand \ \hat{\cal E}_{1}=\hat{\cal E}_{4}$. Therefore, (according to theorem 2) we predict having time-independent pointer states for the system given by $|\pm\rangle=\frac{1}{\sqrt{2}}\{|a\rangle\pm e^{-i\varphi}|b\rangle\}$ QED.

We note that theorem 4 definitely is not contained in theorem 3 and generally these two theorems refer to different conditions for having time-independent pointer states. This is because the condition for having time-independent pointer states represented in equation (\ref{80}) of theorem 4 requires having $\hat{\cal E}_{2}=\hat{\cal E}_{3}e^{-i\varphi}$ and $\hat{\cal E}_{1}=\hat{\cal E}_{4}$ as we discussed. However, according to equation (\ref{71}) satisfaction of this latter condition necessarily will not lead to satisfaction of the condition $[\hat{H}_{\cal S},\hat{H}^{'}]=0$ of theorem 3; unless $\hat{\cal E}_{2}=\hat{\cal E}_{3}=0$ or $s_{1}=s_{4}$ as well. In other words if the condition of theorem 4 (equation (\ref{80})) is satisfied but none of the conditions $\hat{\cal E}_{2}=\hat{\cal E}_{3}=0$ or $s_{1}=s_{4}$ are satisfied, then the condition for commutativity of $\hat{H}_{\cal S}$ and $\hat{H}^{'}$ will not be satisfied.

\section{Conclusion}
Defining the pointer states of a subsystem as those states which are characterized by their ability not to entangle with the states of another subsystem, we presented a general method for evaluating the pointer states of a subsystem. This way we showed how in practice the global state of the system and the environment may evolve into a \emph{diagonal} state (i.e.\ the Von Neumann scheme of measurement may be realized) as a result of the natural evolution of the total composite system. As we showed, evaluation of the pointer states of the system requires finding those specific initial states of the system and the environment for which the operator $\hat{G}(t)$ (defined through equation (\ref{30})) may become independent of the states of the environment. However, as we could see from our example represented by the evolution of the two-level atom in the Jaynes-Cummings model and in the exact-resonance-regime, such initial conditions for the states of the system and the environment necessarily do not exist in an arbitrary regime. (For this example as we saw, unless we have a large average number of photons which can make a sharp distribution function for the state of the electromagnetic field, the states of the two-level atom and the field will remain highly entangled and the pointer states of measurement cannot be realized at all.) As a result, even time-dependent pointer states necessarily do not exist for any arbitrary regime. In other words, premeasurement by the environment (the Von Neumann scheme of measurement), as a result of which the state of the total composite system becomes in a diagonal form, necessarily cannot be realized in any arbitrary regime.

In this paper we distinguished between pointer states of measurement and the preferred basis of measurement; as time-independent pointer states which can arise only in certain regimes. We exactly showed why indeed time-independent pointer states (which require the operator $\hat{G}(t)$ defined through equation (\ref{30}) be independent of time) cannot be expected in most of the regimes. In other words, pointer states of a system, which do not entangle with the states of the environment and appear as a result of premeasurement by the environment, necessarily are not time-independent and the assumption of having quantum \emph{nondemolition premeasurement} in the Von Neumann scheme of measurement practically is not a good assumption; as the pointer states which appear on the diagonal state of the total composite system may change by more than just an overall phase factor. Moreover, we explored those conditions under which the pointer states of the system may be independent from time; so that they can represent the preferred basis of measurement. These are new aspects not contained in the existing theory for determination of the preferred basis of measurement.

As we saw the conditions for having time-independent pointer states include the so-called quantum limit of decoherence ($\hat{H}\approx \hat{H}_{\cal S}$) as well as the so-called quantum measurement limit ($\hat{H}\approx \hat{H}_{\rm int}$). In fact, time-independent pointer states for the system are predicted for these two regimes by using theorem 3 and just as special cases of the more general symmetry condition represented by this theorem. Therefore, our theorems cover the predictions of Zurek's theory for determination of the preferred basis of measurement at corresponding limits. Nonetheless, they present some other conditions as well under which the pointer states of the system would become independent of time and hence can we have a preferred basis of measurement. For example as we saw, in order to have the preferred basis of measurement given by the eigenstates of the interaction Hamiltonian in the Hilbert space of the system, necessarily we do not require $\hat{H}_{\cal S}$ and $\hat{H}_{\cal E}$ to be negligible and this prediction holds valid whenever $\hat{H}_{\cal S}$ commutes with a nonzero $\hat{H}_{\rm int}$ and no matter how big are the contributions from the self-Hamiltonian of the system and the self-Hamiltonian of the environment. Therefore, our criteria for predicting the time-independence of pointer states go beyond the limits in which $\hat{H}_{\rm tot}\approx\hat{H}_{\rm int}$ or $\hat{H}_{\rm tot}\approx\hat{H}_{\cal S}$ and will include some other cases as well; where all contributions can be present at the same time. In this sense, our theory not only provides us with a general method for obtaining the generally time-dependent pointer states of the system and the environment, but also it can serve as a generalization for the existing theory for determination of the preferred basis of measurement.

As an application of our theory, one can use it in order to obtain those regimes of the parameter space (corresponding to the total Hamiltonian defining a given system-environment model) for which a preferred basis of measurement can be realized. Moreover, we can predict the corresponding preferred basis of measurement for each regime. In addition to that now we also have a method in order to obtain time-dependent pointer states in non-measurement regimes; where a time-independent basis of measurement cannot be realized at all. This ability to obtain \emph{time-dependent} pointer states, which arise in the majority of regimes, is particularly important in decoherence studies; as such pointer states although evolve with time and cannot represent the preferred basis of measurement, they correspond to the initial conditions for the state of the system and the environment for which we can have long decoherence times. We will present some very interesting results regarding this problem in our other paper \cite{paper3} where we obtain the time-dependent pointer states of the generalized spin-boson model and study the decoherence of the central system in this model.



\end{document}